\documentclass[nofootinbib, superscriptaddress]{revtex4-2}
\usepackage[utf8]{inputenc}
\usepackage{bm, amsthm, amsmath, amsfonts, amssymb, color, graphicx, natbib}
\usepackage{enumitem}
\usepackage{slashed}
\usepackage{xcolor}
\usepackage{verbatim}
\usepackage{float}
\usepackage{booktabs}
\usepackage{supertabular}
\usepackage{longtable}
\usepackage{array}
\graphicspath{{fig/}}
\usepackage{hyperref}
\hypersetup{
  colorlinks=true,
  linkcolor=blue,
  filecolor=blue,
  urlcolor=blue,
  citecolor=blue
}

\begin{document}

\title{Reconstruction of Primordial Power Spectrum from Gravitational Waves of High-Redshift Black Hole Binaries}
\author{Qianhang Ding}
\email{dingqh@ibs.re.kr}
\affiliation{Cosmology, Gravity and Astroparticle Physics Group, Center for Theoretical Physics of the Universe,
Institute for Basic Science (IBS), Daejeon, 34126, Korea}
\author{Xinpeng Wang}
\email{xinpeng.wang@ipmu.jp}
\affiliation{Kavli Institute for the Physics and Mathematics of the Universe (WPI),
The University of Tokyo Institutes for Advanced Study,
The University of Tokyo, Chiba 277-8583, Japan}
\affiliation{Department of Physics, Graduate School of Science,
The University of Tokyo, Tokyo 113-0033, Japan}
\affiliation{School of Physics Science and Engineering, Tongji University, Shanghai 200092, China}
\author{Masahide Yamaguchi}
\email{gucci@ibs.re.kr}
\affiliation{Cosmology, Gravity and Astroparticle Physics Group, Center for Theoretical Physics of the Universe,
Institute for Basic Science (IBS), Daejeon, 34126, Korea}
\affiliation{Department of Physics, Institute of Science Tokyo,
2-12-1 Ookayama, Meguro-ku, Tokyo 152-8551, Japan}
\affiliation{Department of Physics and IPAP, Yonsei University, 50 Yonsei-ro, Seodaemun-gu, Seoul 03722, Korea}
\author{Ying-li Zhang}
\email{yingli@tongji.edu.cn}
\affiliation{School of Physics Science and Engineering, Tongji University, Shanghai 200092, China}
\affiliation{Institute for Advanced Study of Tongji University, Shanghai 200092, China}
\affiliation{Kavli Institute for the Physics and Mathematics of the Universe (WPI),
The University of Tokyo Institutes for Advanced Study,
The University of Tokyo, Chiba 277-8583, Japan}
\affiliation{Asia Pacific Center for Theoretical Physics, Pohang 37673, Korea}
\affiliation{Center for Gravitation and Cosmology, Yangzhou University, Yangzhou 225009, China}

\begin{abstract}
High-redshift binary black hole (BBH) events are promising candidates for primordial black holes (PBHs) detectable by next-generation gravitational wave (GW) detectors. A redshifted mass distribution of detected PBH candidates can be obtained from GW observations, from which the underlying PBH mass function can be reconstructed. In this work, we develop a framework that applies the gradient-descent method to the observed redshifted mass distribution and reconstructs the PBH mass function and, subsequently, the primordial power spectrum (PPS) on small scales. As an illustrative application, we analyze BBH events in the LIGO–Virgo–KAGRA (LVK) catalogs under a specified PBH selection criterion. We find a regularization-stable candidate bump-like enhancement of order $\mathcal{O}(10^{-2})$ in the reconstructed PPS, centered around $k_{\mathrm{peak}}\simeq 5.7\times 10^5~\mathrm{Mpc}^{-1}$ under the adopted assumptions. Our results demonstrate the feasibility of reconstructing the small-scale PPS from high-redshift BBH observations with next-generation GW detectors.
\end{abstract}

\maketitle

\section{Introduction}

Primordial black holes (PBHs) are hypothetical compact objects that may have formed in the early Universe through the gravitational collapse of highly overdense regions seeded by primordial perturbations. A well-known formation mechanism involves an enhancement of the small-scale primordial power spectrum (PPS), which can generate density perturbations large enough to exceed the collapse threshold upon horizon re-entry~\cite{Hawking:1971ei, Zeldovich:1967lct, Carr:1974nx, Meszaros:1974tb, Carr:1975qj}. Since the PBH population is determined by the primordial curvature perturbations, extracting the PBH mass function from astrophysical observations offers a potential probe of the PPS on small scales~\cite{Kimura:2021sqz, Wang:2022nml}.

Gravitational waves (GWs) provide a unique channel to study the nature of PBHs. Since the detection of the first GW event, GW150914, from a binary black hole (BBH) merger \cite{LIGOScientific:2016aoc}, a number of BBH events suggest the existence of PBHs. For instance, GW190425 and GW190814 have a companion mass smaller than $3 \, M_\odot$ \cite{LIGOScientific:2020aai, LIGOScientific:2020zkf} that can be PBH candidates \cite{Clesse:2020ghq}, GW190521 and GW231123 report the BH companion mass in the astrophysical BH mass gap \cite{LIGOScientific:2020iuh, LIGOScientific:2025rsn} that hints at the potential primordial origin \cite{Clesse:2020ghq,DeLuca:2020sae, Yuan:2025avq}. To ensure the detection of PBHs, high-redshift BBH events are promising candidates \cite{Ding:2020ykt}. The next-generation GW detectors, such as Einstein Telescope \cite{Punturo:2010zz}, DECIGO \cite{Kawamura:2011zz}, LISA \cite{LISA:2017pwj}, and TianQin/Taiji \cite{TianQin:2015yph, Hu:2017mde}, are expected to detect BBHs from redshift $z > 20$ \cite{Klein:2015hvg, Nakamura:2016hna, Ng:2022agi, Wang:2019ryf, Ruan:2018tsw}, where the contamination from astrophysical black holes is expected to be strongly suppressed.

In this work, we propose a novel population-level inverse framework that turns the redshifted mass distribution of high-redshift BBH events into the PBH mass function and, subsequently, the small-scale PPS.
For detected high-redshift BBH events, GW observations provide measurements of the redshifted BH masses, which encode information about both the intrinsic PBH masses and their redshifts~\cite{LIGOScientific:2016vlm}. Given a sufficiently large sample of PBH binary events, one can reconstruct the distribution of the redshifted PBH mass.  This distribution depends on the underlying PBH mass function, the redshift distribution of PBH binaries, and the detector sensitivity.  To infer the PBH mass function from the redshifted mass distribution, we start from an initial PBH mass function and compute the corresponding predicted redshifted mass distribution. We then apply a gradient-descent method~\cite{2016arXiv160904747R} to minimize the discrepancy between the predicted and observed redshifted mass distributions, thereby reconstructing the PBH mass function~\cite{Ding:2023smy}. The PBH abundance is subsequently obtained by integrating the reconstructed mass function over the PBH mass. Finally, to connect the PBH abundance with the PPS, we employ the Press-Schechter formalism and solve the corresponding inverse problem using Tikhonov regularization, enabling reconstruction of the PPS on small scales.

For illustration, we apply this reconstruction method to the BBH events in LIGO-Virgo-KAGRA (LVK) catalogs \cite{LIGOScientific:2018mvr, LIGOScientific:2020ibl, KAGRA:2021vkt, LIGOScientific:2025slb,  LIGOScientific:2026sit}. To select potential PBH candidates in LVK catalogs, we consider two BH mass ranges as selection criteria of PBH candidates: $m_{\rm BH}>15 \,M_\odot$ and $m_{\rm BH}>50\,M_\odot$. The former range is expected to be less dominated by mergers of Pop I and Pop II BHs~\cite{Tanikawa:2021qqi}, while the latter lies in the astrophysical BH mass gap associated with pair-instability supernovae~\cite{2019ApJ...887...53F, Farmer:2020xne, Woosley:2021xba, Tong:2025wpz}. Reconstructing the PBH mass function under these two selection criteria provides a consistency check on the inferred results. After obtaining the PBH mass function, the corresponding PPS is numerically reconstructed by solving the inverse problem of the Press-Schechter formalism with Tikhonov regularization. The peak feature in the reconstructed PBH mass function then maps to a bump-like enhancement in the PPS on small scales. 
We understand that it is still challenging to identify PBHs in LVK catalogs. So this part is only for illustration of our reconstruction framework. However, once we obtain BBH data from much higher redshifts, where astrophysical BBHs are expected to be absent, high-redshift events would greatly reduce astrophysical contamination, making the PBH interpretation cleaner, and this ensures that our method would work better. 

This paper is organized as follows. In Sec.~\ref{sec:roadmap}, we provide the roadmap of reconstruction from the observation to PPS. In Sec.~\ref{sec:redshifted_mass_dist}, we connect the observed redshifted mass distribution with PBH mass function. In Sec.~\ref{sec:PBH_mass_reconst}, we apply the gradient-descent method to reconstruct PBH mass function from the redshifted mass distribution. In Sec.~\ref{sec:reconstruction_LVK}, we use BBH events in LVK to reconstruct potential PBH mass function. In Sec.~\ref{sec:PPS_reconstruction}, we reconstruct PPS on small scale and also considered the effect of non-Gaussianity in the density contrast field. In Sec.~\ref{sec:conclusion}, we conclude our results.

\section{Roadmap}\label{sec:roadmap}
Since the detection of the first GW event, GW150914, GWs have become a powerful tool to study the BH properties, such as BH mass function \cite{Bouhaddouti:2024ena, Tong:2025wpz}. To reveal the nature of BHs from BH mass function, the redshift information should also be involved to take the formation history of BBHs into account in extracting BH mass function, which is crucial to study properties of high-redshift BHs, especially extracting PBH mass function from the future GW observations. We summarize the logic of our reconstruction method as follows. GW observations provide the redshifted component masses $m_z=(1+z)m$, from which one can construct the observed redshifted mass distribution $P_{O} (m^{z}_{1},m^z_{2})$. This distribution is related to the intrinsic PBH mass function $f(m)$, after accounting for the PBH binary redshift distribution $p(z)$, the detector window function $W(m_1,m_2;z)$, and the binary formation mass weighting factor $\eta(m_1,m_2)$. We therefore first solve the inverse problem $P_{O} (m^{z}_{1},m^z_{2})\rightarrow f(m)$ using gradient-descent method. The reconstructed mass function, together with the observed merger abundance determines the total PBH fraction $f_{\mathrm{PBH}}$. Finally, using Press-Schechter relation, $f_{\mathrm{PBH}}(m)$ is converted into the collapse fraction $\beta(m)$, then into the variance of the smoothed density contrast $\sigma^2(R)$, and ultimately into the primordial curvature power spectrum $\mathcal{P}_\mathcal{R}(k)$ through a regularized inverse convolution. In this way, the full reconstruction chain is
\begin{equation}\label{eq:roadmap}
    (m^{z}_{1},m^z_{2})_{\mathrm{obs}}\Rightarrow P_{O} (m^{z}_{1},m^z_{2})\Rightarrow f(m)\Rightarrow f_{\mathrm{PBH}} (m)\Rightarrow\beta(m)\Rightarrow\sigma^2(R)\Rightarrow\mathcal{P}_\mathcal{R}(k).
\end{equation}

\section{Redshifted mass distribution of observed PBH binaries at high redshifts} \label{sec:redshifted_mass_dist}

 In the following discussion, we focus on developing a formalism that connects the observed redshifted mass distribution of PBH binaries with the PBH mass function, and use this formalism to extract PBH mass function from the GW observations. To simplify our formalism, we adopt following three assumptions,
\begin{itemize}[itemsep=1pt,topsep=5pt,parsep=1pt]
	\item[1.] The initial PBH spatial distribution follows the Poisson distribution.
	\item[2.] PBH binaries decouple from the background Hubble flow at the same cosmic time $t_{\rm dec}$ which is early enough.
	\item[3.] The initial PBH mass function has negligible late time evolution.
\end{itemize}

From GW waveform of PBH binaries, we can extract the redshifted mass of PBHs $m_z$, which is a combination of their redshift and intrinsic mass follows $m_z = (1+z)m$ \cite{Rosado:2015voo}. With the development of the next-generation GW detectors, such as Einstein Telescope \cite{Punturo:2010zz}, LISA \cite{LISA:2017pwj}, DECIGO \cite{Kawamura:2011zz}, etc., they are able to detect high-redshift BBH events with a redshift $z > 20$, and this promises the primordial origin of BBH events \cite{Ding:2020ykt, Nakamura:2016hna}. For each high-redshift BBH event, a redshifted mass pair $(m_1^z,\,m_2^z)$ can be obtained from two BH components \footnote{We use $z$ as superscript in $m_i^z$ to denote the redshifted mass of $i$th binary component, while $z$ is subscript in $m_z$ for referring the redshifted BH mass in general. }. After detecting enough high-redshift BBH events, a redshifted mass distribution of PBH binaries $P(m_1^{z}, m_2^{z})$ can be statistically constructed, which is defined as follows,
\begin{align}\label{eq:redshifted_mass_dist}
	P(m_1^z, m_2^z) = \frac{1}{N_\mathrm{tot}} \frac{dN_\mathrm{obs}(m_1^z, m_2^z)}{dm_1^z \, dm_2^z}~,
\end{align}
where $dN_\mathrm{obs}(m_1^z, m_2^z)$ is the number of observed PBH binaries within redshifted mass range $(m_1^z, m_1^z + dm_1^z)$ and $(m_2^z, m_2^z + dm_2^z)$ on the past light cone, $N_\mathrm{tot}$ is the total number of PBH binaries on the past light cone, $m_1^z$ and $m_2^z$ are the redshifted mass of each PBH component in binaries. Here, we should notice that the redshifted mass distribution of PBH binaries $P(m_1^{z}, m_2^{z})$ is not normalized, which means $N_{\rm tot}>\int \int \frac{dN_{\rm obs}}{d m_1^z dm_2^z} dm_1^z dm_2^z = N_{\rm obs}$.

Intuitively, the observed redshifted mass distribution of PBH binaries depends on PBH mass function, sensitivity of GW detectors, and redshift distribution of PBH binaries. To find this relation, we start from calculating the cumulative distribution of redshifted PBH mass $C(m_1^z, m_2^z)$, and its definition and relation with $P(m_1^z, m_2^z)$ can be described as follows,
\begin{align}\label{eq:cumulative_dist}
	C(m_1^z, m_2^z) = \frac{N_\mathrm{obs}(m_a < m_1^z, m_b < m_2^z)}{N_\mathrm{tot}} = \int_0^{m_1^z} \int_0^{m_2^z} P(m_a, m_b) dm_a dm_b~,
\end{align}
where $N_\mathrm{obs}(m_a < m_1^z, m_b < m_2^z)$ is the number of observed PBH binaries with the redshifted mass of their components smaller than $m_1^z$ and $m_2^z$. 

To express $C(m_1^z, m_2^z)$ in the form of PBH mass function $f(m)$, we first define PBH mass function $f(m)$ as the quantity $f(m) dm$ is the probability that a randomly chosen PBH lies in the mass range $(m, m+dm)$, it can be expressed in the form of the PBH number density distribution as follows,
\begin{align}\label{eq:pbh_massfun}
    f(m) = \frac{1}{n_\mathrm{PBH}} \frac{dn}{dm}~,
\end{align}
where $dn$ denotes the comoving number density of PBHs in the mass range $(m, m+dm)$, and $n_\mathrm{PBH}$ is the comoving number density of all PBHs, which can be calculated as $n_\mathrm{PBH} = \int dn$. Therefore PBH mass function is normalized as $\int f(m) dm = 1$. Then we quantify the probability of PBH binaries with intrinsic mass in the range of $(m_1, m_1+dm_1)$ and $(m_2, m_2+dm_2)$ among all PBH binaries, we can start from calculating the number density of PBH binaries with intrinsic mass in the range of $(m_1, m_1+dm_1)$ and $(m_2, m_2+dm_2)$. 

\begin{figure}[htbp]
	\centering
	\includegraphics[width=8cm]{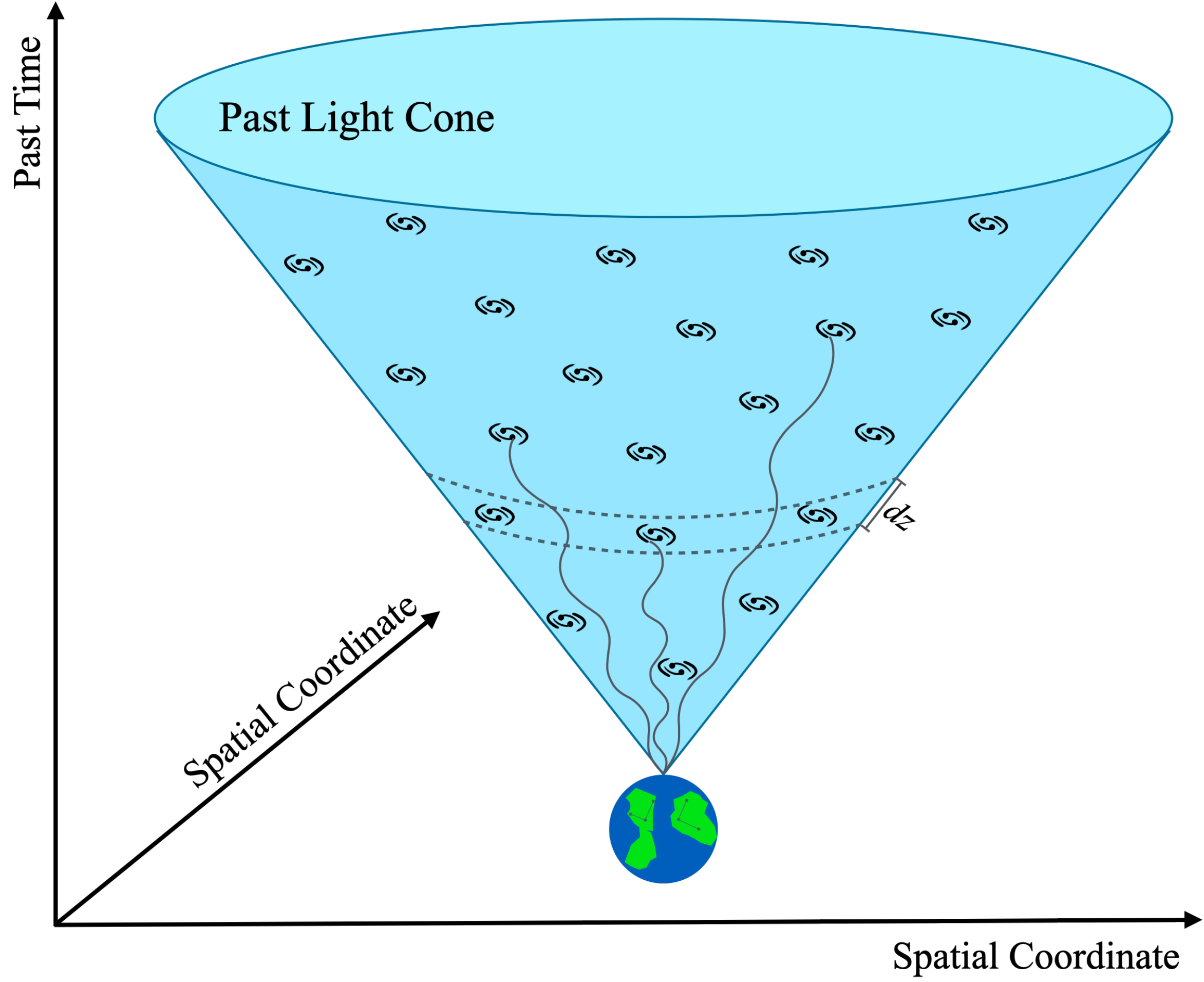}
	\caption{The distribution of PBH binaries on the past light cone. Their emitted GW can propagate along the line of sight from PBH binaries to LVK detectors. PBH binaries within the redshift interval $(z, z+dz)$ are shown between two gray dashed lines. Counting the number of these PBH binaries at various redshift intervals and dividing it by the total number of PBH binaries on the past light cone can produce a redshift distribution $p(z)$ of PBH binaries in Eq.~\eqref{eq:redshift_distribution}.}
	\label{fig:PBH_redshift_dependence}
\end{figure}

Consider a spherical volume with a radius of $r$, there are only two PBHs in this spherical volume and their comoving separation is $x$. 
Based on Poisson distribution in Eq.~(2.18) of \cite{Raidal:2018bbj}, the expected comoving number density of PBH binaries with mass $m_1$ and $m_2$ satisfies
\begin{align}\label{eq:dnb}
	dn_b(m_1, m_2) = \frac{1}{2} e^{-\bar{N}(r)} n_\mathrm{PBH} f(m_1) dm_1 n_\mathrm{PBH} f(m_2) dm_2 V(x_{\rm dec})~,
\end{align}
where $\bar{N}(r) \equiv n_{\rm PBH} V(r) = \frac{4}{3} \pi r^3 n_{\rm PBH}$ is the expected number of PBH in this spherical volume. $V(x_{\rm dec})$ is the comoving volume that PBH binaries decouple from Hubble-flow, which satisfies $\frac{m_1+m_2}{a^3(t_{\rm dec}) V(x)} > \rho_{\rm bg}$ for any PBH separation $x < x_{\rm dec}$, $a(t_{\rm dec})$ is the scale factor at PBH decouple time $t_{\rm dec}$, and $\rho_{\rm bg}$ is the background cosmic density at $t_{\rm dec}$. In this comoving volume $V(x_{\rm dec})$, there are only two PBHs inside to form a binary, and more PBHs inside would disrupt the binary system \cite{Raidal:2018bbj}.
Then we can estimate the total comoving number density of PBH binaries $n_b$ as
\begin{align}
	n_b = \int dn_b(m_1, m_2) = \frac{1}{2} e^{-\bar{N}(r)} n_\mathrm{PBH}^2  \int_0^\infty \int_0^\infty f(m_1) f(m_2) V(x_{\rm dec}) dm_1 dm_2~.
\end{align}
Here, we assume PBH binaries formed at the same cosmic time that corresponds with the same $\rho_{\rm bg}$, and it leads to $V(x_{\rm dec}) \propto m_1 + m_2$, and we ignore the weak dependence of $\bar{N}(r)$ on the mass of PBH binary. Then, we have the probability of PBH binary number density within mass range $(m_1, m_1+dm_1)$ and $(m_2, m_2+dm_2)$ among all PBH binary number density in this spherical volume,
\begin{align}\label{eq:pbh_binary_fraction}
	\frac{dn_b(m_1, m_2)}{n_b} = f(m_1) f(m_2) \eta(m_1,m_2) dm_1 dm_2~,
\end{align}
where $\eta(m_1,m_2)$ is defined as
\begin{align}
   \eta(m_1,m_2)= \frac{m_1+m_2}{\iint (m_1+m_2)f(m_1) f(m_2) dm_1 dm_2}~.
\end{align}
According to this PBH binary number density probability, we can extend it from this spherical volume to the whole observed universe, that gives the probability of the number of PBH binaries within mass range $(m_1, m_1+dm_1)$ and $(m_2, m_2+dm_2)$ in all PBH binaries on the past light cone $dN(m_1,m_2)/N_{\rm tot} \propto dn_b(m_1, m_2)/n_b$. This produces the PBH binary distribution density as
\begin{align}\label{eq:pbh_mass_dist}
    \frac{1}{N_{\rm tot}}\frac{dN(m_1,m_2)}{ dm_1 dm_2} = f(m_1) f(m_2) \eta(m_1, m_2)~.
\end{align}
Since, we are interested in the redshifted mass distribution of PBH binaries $P(m_1^z, m_2^z)$, it is necessary to introduce the redshift dependence of PBH binaries.
In the observation, GWs from PBH binaries propagate along the line of sight, and we can use the number distribution of PBH binaries on the past light cone to estimate its redshift distribution $p(z)$, which is defined as
\begin{align}\label{eq:redshift_distribution}
    p(z) = \frac{1}{N_z(m_1,m_2)} \frac{dN_z (m_1,m_2)}{dz}~,
\end{align}
where $dN_z(m_1,m_2)$ is the number of PBH binaries in the mass range $(m_1, m_1+dm_1)$ and $(m_2, m_2+dm_2)$ along the line of sight within the redshift range $(z, z+dz)$, and $N_z(m_1,m_2) = \int dN_z(m_1,m_2)$ is the total number of PBH binaries within the mass range $(m_1, m_1+dm_1)$ and $(m_2, m_2+dm_2)$ on the past light cone, which is the same as $dN(m_1,m_2)$ in Eq.~\eqref{eq:pbh_mass_dist}. In principle, the redshift distribution of PBH binaries should be BH mass dependent $p(z) \equiv p(z | m_1, m_2)$. For PBH binary distribution on the past light cone, their comoving number density is almost constant (ignoring PBH binary mergers or disruption), then $dN_z(m_1,m_2) = n_b(m_1,m_2) dV_c$ and this causes the mass dependence term $n_b(m_1,m_2)$ in $dN(m_1,m_2)$ and $N_z(m_1,m_2)$ cancel with each other and gives a mass-independent redshift distribution $p(z)$.
We also show the PBH binaries on the past light cone in Fig.~\ref{fig:PBH_redshift_dependence} to intuitively illustrate the redshift dependence of PBH binaries.
Then we combine $p(z)$ with Eq.~\eqref{eq:pbh_mass_dist} to obtain PBH binary distribution density with respect to $(m_1,m_2,z)$ as
\begin{align}
    \frac{1}{N_{\rm tot}}\frac{dN_z(m_1,m_2)}{ dm_1 dm_2 dz} = f(m_1) f(m_2) \eta(m_1, m_2) p(z)~.
\end{align}

However, not all the PBH binaries with intrinsic mass $m_1$ and $m_2$ can be detected due to the limited sensitivity of GW detectors, it depends on their orbital parameters and corresponding redshift, so we use detectable window function $W(m_1, m_2; z)$ to give the detection probability of the observed PBH binaries among all existing PBH binaries with intrinsic mass $m_1$ and $m_2$ at redshift $z$, which is defined as,
\begin{align}\label{eq:window_function}
    W(m_1, m_2; z) = \frac{dN_\mathrm{obs}^z(m_1, m_2)}{dN_z(m_1, m_2)}~,
\end{align}
where $dN_\mathrm{obs}^z(m_1, m_2)$ denotes the observed number of PBH binaries within intrinsic mass range $(m_1, m_1+dm_1)$ and $(m_2, m_2+dm_2)$ in redshift interval $(z, z+dz)$ on the past light cone. 

These observed PBH binaries $dN_\mathrm{obs}^z(m_1, m_2)$ contribute to the amount of PBH binaries with redshifted mass $m_1^z = m_1 (1+z)$ and $m_2^z = m_2 (1+z)$ in the observation of redshifted mass distribution of PBH binaries, and we can use it to calculate the cumulative distribution $C(m_1^z, m_2^z)$ as
\begin{align}
    \frac{dC(m_1^z, m_2^z)}{dm_1 dm_2 dz} = \frac{1}{N_{\rm tot}}\frac{dN_{\rm obs}^z (m_1,m_2)}{dm_1 dm_2 dz}= f(m_1) f(m_2) \eta(m_1, m_2) p(z) W(m_1, m_2; z)~.
\end{align}
To evaluate $C(m_1^z, m_2^z)$, we first calculate its contribution within redshift range $(z, z+dz)$. For a given redshift $z$, the counted PBH binaries should satisfy that their intrinsic mass are in the range of $(0, m_1^z/1+z)$ and $(0, m_2^z/1+z)$, then we integrate the redshifted mass contribution over this intrinsic mass range, which can be expressed as follows,
\begin{align}
	\frac{dC(m_1^z, m_2^z)}{dz} = \int_0^{\frac{m_1^z}{1+z}} \int_0^{\frac{m_2^z}{1+z}} f(m_1) f(m_2) \eta(m_1, m_2) W(m_1, m_2; z) p(z) dm_1 dm_2~.
\end{align}
Then we accumulate all the counted PBH binaries from different redshifts to obtain $C(m_1^z, m_2^z)$ as
\begin{align}\label{eq:cumulative}
	C(m_1^z, m_2^z) = \int_0^\infty \int_0^{\frac{m_1^z}{1+z}} \int_0^{\frac{m_2^z}{1+z}} f(m_1) f(m_2) \eta(m_1, m_2) W(m_1, m_2; z) p(z) dm_1 dm_2 dz~.
\end{align}
Here we formally integrate redshift in the range of $(0, \infty)$ and $(0, m_z/1+z)$. However, in practice, the bound of integration should be in a physical range like $(m_\mathrm{min}, m_z/1+z)$ which is determined by PBH mass distribution, and $(z_\mathrm{min}, z_\mathrm{max})$ which is determined by the GW detector. After obtaining $C(m_1^z, m_2^z)$, the redshifted mass distribution $P(m_1^z, m_2^z)$ can be calculated by deriving the cumulative distribution with respect to the redshifted mass as 
\begin{align}
P(m_1^z, m_2^z) = \frac{\partial^2 C(m_1^z, m_2^z)}{\partial m_1^z \partial m_2^z}~.
\end{align}
We first take the partial derivative of $C(m_1^z, m_2^z)$ with respect to $m_2^z$ as follows,
\begin{align}\label{appeq:cumu_1}\nonumber
	\frac{\partial C(m_1^z, m_2^z)}{\partial m_2^z} &= \lim_{\Delta m_2^z \to 0} \frac{1}{\Delta m_2^z} \int_0^\infty p(z) dz \int_0^{\frac{m_1^z}{1+z}} f(m_1) dm_1 \\\nonumber
	&\times \left(\int_0^{\frac{m_2^z+\Delta m_2^z}{1+z}} f(m_2) \eta(m_1, m_2) W(m_1, m_2; z)dm_2 - \int_0^{\frac{m_2^z}{1+z}} f(m_2) \eta(m_1, m_2) W(m_1, m_2; z)dm_2 \right)\\\nonumber
	&=\lim_{\Delta m_2^z \to 0} \frac{1}{\Delta m_2^z} \int_0^\infty p(z) dz \int_0^{\frac{m_1^z}{1+z}} f(m_1) dm_1 \int_{\frac{m_2^z}{1+z}}^{\frac{m_2^z+\Delta m_2^z}{1+z}} f(m_2) \eta(m_1, m_2) W(m_1, m_2; z)dm_2\\\nonumber
	&=\int_0^\infty p(z) dz \int_0^{\frac{m_1^z}{1+z}} f(m_1) dm_1 \times \lim_{\Delta m_2^z \to 0} \frac{1}{\Delta m_2^z} \int_{\frac{m_2^z}{1+z}}^{\frac{m_2^z+\Delta m_2^z}{1+z}} f(m_2)\eta(m_1, m_2) W(m_1, m_2; z)dm_2\\
	&=\int_0^\infty \int_0^{\frac{m_1^z}{1+z}} f(m_1) f\left(\frac{m_2^z}{1+z}\right) \eta\left(m_1,\frac{m_2^z}{1+z}\right) W\left(m_1, \frac{m_2^z}{1+z}; z\right) \frac{p(z)}{1+z} dm_1 dz~.
\end{align}
Then we take the partial derivative of Eq~\eqref{appeq:cumu_1} with respect to $m_1^z$ as follows,
\begin{align}\label{appeq:cumu_2}\nonumber
	&\frac{\partial}{\partial m_1^z} \left(\frac{\partial C(m_1^z, m_2^z)}{\partial m_2^z}\right) \\\nonumber
    &= \lim_{\Delta m_1^z \to 0} \frac{1}{\Delta m_1^z} \int_0^\infty f\left(\frac{m_2^z}{1+z}\right) \frac{p(z)}{1+z} dz\\\nonumber
	&\times \left(\int_0^{\frac{m_1^z+\Delta m_1^z}{1+z}} f(m_1) \eta\left(m_1,\frac{m_2^z}{1+z}\right) W\left(m_1, \frac{m_2^z}{1+z}; z\right) dm_1 - \int_0^{\frac{m_1^z}{1+z}} f(m_1) \eta\left(m_1,\frac{m_2^z}{1+z}\right) W\left(m_1, \frac{m_2^z}{1+z}; z\right) dm_1 \right)\\\nonumber
	&=\lim_{\Delta m_1^z \to 0} \frac{1}{\Delta m_1^z} \int_0^\infty f\left(\frac{m_2^z}{1+z}\right) \frac{p(z)}{1+z} dz \int_{\frac{m_1^z}{1+z}}^{\frac{m_1^z+\Delta m_1^z}{1+z}} f(m_1) \eta\left(m_1,\frac{m_2^z}{1+z}\right) W\left(m_1, \frac{m_2^z}{1+z}; z\right) dm_1\\\nonumber
	&=\int_0^\infty f\left(\frac{m_2^z}{1+z}\right) \frac{p(z)}{1+z} dz \times \lim_{\Delta m_1^z \to 0} \frac{1}{\Delta m_1^z} \int_{\frac{m_1^z}{1+z}}^{\frac{m_1^z+\Delta m_1^z}{1+z}} f(m_1) \eta\left(m_1,\frac{m_2^z}{1+z}\right) W\left(m_1, \frac{m_2^z}{1+z}; z\right) dm_1\\
	&= \int_0^\infty f\left(\frac{m_1^z}{1+z}\right) f\left(\frac{m_2^z}{1+z}\right) \eta\left(\frac{m_1^z}{1+z}, \frac{m_2^z}{1+z}\right) W\left(\frac{m_1^z}{1+z}, \frac{m_2^z}{1+z}; z\right) \frac{p(z)}{(1+z)^2} \, dz~.
\end{align}
Then we obtain the expression of $P(m_1^z, m_2^z)$ as follows,
\begin{align}\label{eq:probability}
	P(m_1^z, m_2^z) &= \frac{\partial^2 C(m_1^z, m_2^z)}{\partial m_1^z \, \partial m_2^z}\nonumber\\ 
	&= \int_0^\infty f\left(\frac{m_1^z}{1+z}\right) f\left(\frac{m_2^z}{1+z}\right) \eta\left(\frac{m_1^z}{1+z}, \frac{m_2^z}{1+z}\right) W\left(\frac{m_1^z}{1+z}, \frac{m_2^z}{1+z}; z\right) \frac{p(z)}{(1+z)^2} \, dz~.
\end{align}
Eq.~\eqref{eq:probability} provides a direct relation between observed redshifted mass distribution and PBH mass function, this indicates that primordial physics from PBH mass function can be extracted from GW observations. However, there are still some mechanisms that would modify the PBH mass function in the late universe. For instance, PBHs accrete surrounding matter and their masses increase accordingly \cite{DeLuca:2020fpg}; Late PBH binaries can also form via two-body capture or three-body interaction in dark matter halos \cite{Franciolini:2022ewd, Ding:2024mro}; PBH binaries could experience disruption process due to matter inhomogeneities \cite{Raidal:2018bbj}. These effects require detailed analysis on PBH binaries dynamics, we leave these effects for future work. In this work, we use Eq.~\eqref{eq:probability} to do the benchmarking calculation as the first step to approach the PBH mass function reconstruction.

\section{Reconstruct PBH mass function from redshifted mass distribution} \label{sec:PBH_mass_reconst}

As we have discussed above, the relation between redshifted mass distribution of PBH binaries and PBH mass function follows Eq.~\eqref{eq:probability} as
\begin{align}\label{eq:probability_1}
	P_O(m_1^z, m_2^z) = \int_0^\infty f_p\left(\frac{m_1^z}{1+z}\right) f_p\left(\frac{m_2^z}{1+z}\right) \eta\left(\frac{m_1^z}{1+z}, \frac{m_2^z}{1+z}\right) W\left(\frac{m_1^z}{1+z}, \frac{m_2^z}{1+z}; z\right) \frac{p(z)}{(1+z)^2} \, dz~,
\end{align}
where $P_O(m_1^z, m_2^z)$ with subscript $O$ denotes the observed redshifted mass distribution and $f_p(m)$ is the physical PBH mass function. Apart from $f_p(m)$, $P_O(m_1^z, m_2^z)$ only depends on detector-dependent window function $W(m_1, m_2; z)$ and redshift distribution of PBH binaries $p(z)$. Once we know the contribution of $W(m_1, m_2; z)$ and $p(z)$, the physical PBH mass function $f_p(m)$ can be inversely solved from observed redshifted mass distribution $P_O(m_1^z, m_2^z)$. {It should be noticed that in constructing $P_O(m_1^z, m_2^z)$ from data, we first normalize $P_O(m_1^z, m_2^z)$ to facilitate numerical implementation, and this would introduce an artificial overall scaling factor on the physical PBH mass function $f_p(m)$, then we recover the physical distribution of $f_p(m)$ by re-normalizing its reconstructed result.}

Solving PBH mass function in Eq.~\eqref{eq:probability_1} is a typical inverse problem, which is hard to find an analytical solution. Hence, we use the gradient-descent method in Eq.~\eqref{eq:probability_1} to reconstruct PBH mass function from redshifted mass distribution. There are several steps in the gradient-descent method as follows,
\begin{itemize}[itemsep=1pt,topsep=5pt,parsep=1pt]
	\item[1.] Give an initial PBH mass distribution as an initial condition of PBH mass function.
	\item[2.] Calculate theoretical redshifted mass distribution and compare it with observed one to obtain the error function.
	\item[3.] Update PBH mass function by calculating the gradient of error function with respect to the variation of PBH mass function.
	\item[4.] Iterate PBH mass function until finding the minimum of error function, which is a best-fit approximation of physical PBH mass function.
\end{itemize}

In order to reduce the computational cost of this algorithm, we start from discretizing PBH mass function $f(m)$ to a PBH mass distribution vector $\mathbf{f}$, where the component of vector follows $\mathbf{f}_\mathrm{i} = f(m_i)$ and a set of $\{m_i| 1 \leq i \leq N\}$ is $N$ mass points taken from a suitable PBH mass range. Then we provide an initial value of this PBH mass distribution vector $\mathbf{f}$. This initial vector could follow an arbitrary mass distribution, however, choosing a more realistic and reasonable initial mass distribution vector could effectively decrease the time complexity of the gradient-descent method. After giving the initial value of $\mathbf{f}$, an initial PBH mass function can be approximately constructed by interpolating the initial PBH mass distribution vector.

For a given assumed PBH mass function $\tilde{f}(m)$, we can combine it with window function $W(m_1, m_2; z)$ and redshift distribution $p(z)$ to calculate a theoretical redshifted mass distribution $P_T(m_1^z, m_2^z)$ as
\begin{align}\label{eq:theoretical_prob}
    P_T(m_1^z, m_2^z) = \int_0^\infty \tilde{f}\left(\frac{m_1^z}{1+z}\right) \tilde{f}\left(\frac{m_2^z}{1+z}\right) \eta\left(\frac{m_1^z}{1+z}, \frac{m_2^z}{1+z}\right) W\left(\frac{m_1^z}{1+z}, \frac{m_2^z}{1+z}; z\right) \frac{p(z)}{(1+z)^2} \, dz~,
\end{align}
Then we can define an error function $E(\mathbf{f})$ by comparing theoretical distribution with the observed redshifted mass distribution at a selected set of PBH binary mass pairs as follows \cite{2023arXiv230105579C},
\begin{align}\label{eq:error_fun}
	E(\mathbf{f}) \equiv \sqrt{\frac{\sum_{1 \leq i \leq j \leq N } [P_T(m_i^z, m_j^z) - P_O(m_i^z, m_j^z)]^2}{(1+N)N/2}}~,
\end{align}
where $P_{T}(m_1^z, m_2^z)$ and $P_{O}(m_1^z, m_2^z)$ are theoretical and observed redshifted mass distribution, respectively, and $N$ is the number of selected redshifted mass points for $m_1^z$ and $m_2^z$. 

The error function $E(\mathbf{f})$ only depends on the assumed PBH mass distribution vector $\mathbf{f}$, so any further deviation of $\mathbf{f}$ from the physical PBH mass function $f_p(m)$ would enhance the difference between $P_T(m_1^z, m_2^z)$ and $P_O(m_1^z, m_2^z)$ and increase the value of error function. It indicates that, in order to minimize the difference between assumed PBH mass distribution vector and the physical PBH mass function, minimizing the error function $E(\mathbf{f})$ could be an indicator to make $\mathbf{f}$ approach the physical PBH mass function. To minimize the error function $E(\mathbf{f})$, we use the following iteration equation to update $\mathbf{f}$,
\begin{align}\label{eq:recursion}
	f_\mathrm{k+1}(m_i) = f_\mathrm{k}(m_i) - \gamma \frac{\partial E(\mathbf{f_\mathrm{k}})}{\partial f_\mathrm{k}(m_i)}~,
\end{align}
where $\mathbf{f}_\mathrm{k}$ and $f_k(m)$ are the $k \, $th iterated PBH mass vector and PBH mass function, and $f_\mathrm{k}(m_i)$ is its $i \,$th component. $\gamma$ is the learning rate in the gradient-descent method, which describes the length of updated step during each iteration. Meanwhile, the updated PBH mass distribution $f_k(m)$ is forced to satisfy $f_k(m) \geq 0$. With the update of $\mathbf{f}$, error function $E(\mathbf{f})$ is approaching its minimal value during iterations, the gradient of error function $\nabla E(\mathbf{f})$ is approaching zero vector, this causes the updated PBH mass distribution vector $\mathbf{f}_\mathrm{k+1} \approx \mathbf{f}_\mathrm{k}$. In this case, the iterated $\mathbf{f}$ can be viewed as a discretized approximation of the physical PBH mass function $f_p(m)$.

However, this kind of inverse problem is usually ill-posed, which means its solution does not satisfy the existence, uniqueness and continuity. 
In this work, we treat the reconstruction as a discretized inverse problem, and it can be shown that the solution to such a discretized inverse problem is unique (see Appendix~\ref{app:well_posedness} for more details). This can make sure the reconstructed mass function is reliable.

\section{PBH mass function reconstruction in LVK data}\label{sec:reconstruction_LVK}

Since the first detection of binary black hole merger event GW150914, LVK collaboration has detected more than $300$ events over past decade. Among these events, some single events and statistical information indicates that PBH binaries could be hidden in them. Meanwhile, these LVK events cover redshift range $0 < z < 1$, which could provide a first evaluation on reconstructing PBH mass function via above method.

To reconstruct PBH mass function via proposed method, we should evaluate the detector-dependent window function $W(m_1,m_2;z)$ and redshift distribution of PBH binaries $p(z)$ in Eq.~\eqref{eq:probability_1}. The detector-dependent window function $W(m_1,m_2;z)$ is defined in Eq.~\eqref{eq:window_function}, it characterizes the fraction of observed events by LVK among all PBH binaries. We adopt a simplified detection criterion that a binary black hole (BBH)
event is detectable if its optimal signal-to-noise ratio (SNR) exceeds $8$ \cite{LIGOScientific:2016vbw}, given by
\begin{align}\label{eq:SNR}
    \text{SNR} = \sqrt{4 \int_{f_{\rm min}}^{f_{\rm max}} \frac{|\tilde{h}(f)|^2}{S_n(f)} df} > 8~,
\end{align}
where the integration range $[f_{\rm min}, f_{\rm max}]$ is the observational frequency range, where $f_{\rm max} = \min(f_{\rm det, max}, f_{\rm ISCO}/(1+z))$. Here $f_{\rm det,max}$ is the maximal detectable GW frequency of detector and $f_{\rm ISCO}$ is the GW frequency at the innermost stable circular orbit of BBHs. $S_n(f)$ is the noise strain of the detectors \cite{Moore:2014lga}. $\tilde{h}(f)$ is the Fourier transform of the GW strain that can be expressed as \cite{Rosado:2015voo, Droz:1999qx}
\begin{align}\label{eq:GW_strain_Fourier}
    \tilde{h}(f) = \sqrt{\frac{5}{24}} \frac{(G \mathcal{M}_c (1+z))^{5/6}}{\pi^{2/3} c^{3/2} d_L(z)} f^{-7/6}~.
\end{align}
Here, $\mathcal{M}_c \equiv (m_1 m_2)^{3/5}/(m_1 + m_2)^{1/5}$ is the chirp mass of PBH binary. $d_L(z)$ is the luminosity distance between PBH binary and the observer. $G$ is the Newton's gravitational constant and $c$ is the speed of light.

The detected LVK events are the accumulated events over past ten year observations since $2015$, it indicates that the LVK has potential to detect the PBH binaries with a frequency that can evolve to LVK frequency band within ten years. And once a PBH binary reaches LVK frequency band with a $\text{SNR} \geq 8$, this event is detected. 

The key point to estimate the detectable window function is to find the critical GW frequency $f_c$ that can evolve into LVK frequency band during ten-year evolution, and the PBH binaries with a frequency larger than this critical frequency should be detected if their $\text{SNR} \geq 8$. Hence the detectable window function $W(m_1, m_2 ; z)$ at redshift $z$ should be an integration of the probability distribution $P_t$ of PBH binary orbital parameters over the frequency larger than $f_c$ as
\begin{align}\label{eq:windowfun}
    W(m_1, m_2; z) = \int_{f_c}^{f_{\rm max}} P_t(f) \Theta ({\rm SNR}-8)df~.
\end{align}
Here, $P_t(f)$ is the probability distribution of GW frequency at cosmic time $t$, which is corresponding with redshift $z$. $\Theta({\rm SNR} - 8)$ is the Heaviside step function that only counts the BBH events with ${\rm SNR} \geq 8$. This probability distribution is time-dependent, because GW emission causes the orbital shrinkage of PBH binaries. In order to estimate $P_t$, we start from its initial probability distribution. The initial probability distribution of the orbital parameter (semi-major axis $a$ and eccentricity $e$) of PBH binaries $P_{\rm ini}$ follows \cite{Sasaki:2016jop, Chen:2018czv}, 
\begin{align}\label{eq:orbit_para_prob}
    P_{\rm ini}(a, e) = \frac{3}{4} \left(\frac{f_b}{\bar{x}}\right)^{3/2} a^{1/2} \frac{e}{(1-e^2)^{3/2}}~,
\end{align}
where $f_b$ is the energy density fraction of PBH binary with mass $m_1$ and $m_2$ in dark matter, $\bar{x}$ is the mean separation of PBH binaries. Since the template-dependent method in LVK detection can only detect a binary system with an eccentricity $e \simeq 0$ due to a lack of eccentricity template \cite{LIGOScientific:2018mvr, LIGOScientific:2019dag}, meanwhile high-eccentricity events would cause the merger time much shorter than that of circularized-orbit events, and these reasons cause non-zero eccentricity events are not taken into consideration. In such a case, $P_{\rm ini}(a) \sim a^{1/2} e$, and consider a small detectable eccentricity range $0<e<e_{\rm cut} = 0.1$ that leads to $P_{\rm ini}(a) \sim a^{1/2}$. Meanwhile, the orbital distribution evolves because GW emission shrinks the orbits, redistributing binaries toward smaller $a$ (higher $f$), with the closest binaries merging out of the surviving population. This causes the initial probability distribution to evolve into a late-time distribution $P_t$, and the evolution of probability distribution follows
\begin{align}\label{eq:prob_t}
    P_t(a_t) = \frac{dn}{da_t} = \frac{dn}{da_{\rm ini}} \frac{da_{\rm ini}}{da_t} =  P_{\rm ini}(a_{\rm ini})\frac{da_{\rm ini}}{da_t}~,
\end{align}
Where $dn$ is the number density of PBH binary within the semi-major axis interval $(a, a+da)$. The relation of semi-major axis at formation $a_{\rm ini}$ and later time $a_t$ can be obtained from the evolution of semi-major axis as described in Ref.~\cite{Peters:1964zz},
\begin{align}
    \frac{da}{dt} = -\frac{64}{5} \frac{G^3 m_1 m_2 (m_1+m_2)}{c^5 a^3}~.
\end{align}
This gives a relation of semi-major axis between the initial value and a later value during the evolution from initial binary formation time $t_{\rm i}$ to the later time $t$
\begin{align}
    a_{\rm ini}^4 = a_{t}^4 + \delta^4(\Delta t)~,
\end{align}
where $\Delta t \equiv t- t_{\rm i}$ is the evolution time since the PBH binary formation. $\delta^4(\Delta t)$ is defined as $\delta^4(\Delta t) \equiv 256 G^3 m_1 m_2 (m_1+m_2) \Delta t/5 c^5$.
It gives $da_{\rm ini}/da_t = a_t^3/a_{\rm ini}^3$, then we can rewrite Eq.~\eqref{eq:prob_t} in the form of
\begin{align}
    P_t(a_t) =  P_{\rm ini}(a_{\rm ini})\frac{da_{\rm ini}}{da_t} = P_{\rm ini}(a_{\rm ini})\frac{a_t^3}{a_{\rm ini}^3} = P_{\rm ini}((a_t^4 + \delta^4(\Delta t))^{1/4}) \frac{a_t^3}{(a_t^4 + \delta^4(\Delta t))^{3/4}}~.
\end{align}
Since PBH binaries form at very high redshift around matter-radiation equality \cite{Ali-Haimoud:2017rtz}, its corresponding cosmic time $t_{\rm i}$ is much smaller than the cosmic time $t$ at low redshift, $t \gg t_{\rm i}$. Then we can approximate evolution time as $\Delta t \simeq t$. It indicates that the later time probability distribution $P_t$ is dependent on time.
In small semi-axis limit $a_t^4 \ll \delta^4(\Delta t)$, this probability distribution can be approximated as
\begin{align}
    P_t(a_t) \simeq \frac{P_{\rm ini}(\delta(\Delta t))}{\delta^3(\Delta t)} a_t^3~. 
\end{align}
Then we can use Kepler's third law to derive the probability distribution over frequency,
\begin{align}
    P_t(f) = P_t(a) \left|\frac{da}{df}\right| = \frac{2}{3} \frac{[G(m_1+m_2)]^{4/3}}{\pi^{8/3}}\frac{P_{\rm ini}(\delta(\Delta t))} {\delta^3(\Delta t)} f^{-11/3}~.
\end{align}
To estimate the window function $W$, we need to estimate critical frequency $f_c$. The evolution of GW frequency in a circular orbit follows
\begin{align}\label{eq:GW_evolution}
     \frac{d f}{d t} &=\frac{96}{5}\frac{\pi^{8/3}}{c^5} (G\mathcal{M}_c)^{5/3}  f^{11/3}~.
\end{align}
Here, the rest-frame GW frequency $f$ is related to the observed GW frequency $f_o$ by $f = (1+z)f_o$. 
Since the critical frequency $f_c$ should evolve to the lower frequency cutoff (denoted $f_{\rm LVK, min}$) by the LVK search within 10 years. We can use Eq.~\eqref{eq:GW_evolution} to determine the $f_c$, given by
\begin{align}\label{eq:fc}
    f_c^{-8/3} - [(1+z)f_{\rm LVK,min}]^{-8/3} = \frac{256}{5} \frac{(G\mathcal{M}_c)^{5/3} \pi^{8/3} }{c^5} \frac{\Delta T}{1+z}~,
\end{align}
where $(1+z)f_{\rm LVK,min}$ denotes the minimal GW frequency of LVK detectors in the rest frame of PBH binaries. In the numerical analysis we set $f_{\rm LVK,min}$ consistently with LVK sensitivity curves. Since $(1+z)f_{\rm LVK, min}$ is much larger than critical frequency $f_c$ in the estimation, the variance in $f_{\rm LVK, min}$ would not affect the result of window function in Eq.~\eqref{eq:windowfun}. $\Delta T$ is the detection time of LVK since its first running in $2015$ which is set as $10 \, {\rm yrs}$ in following calculation and $1+z$ term comes from the cosmic time dilation. Then we can use Eq.~(\ref{eq:windowfun} -- \ref{eq:fc}) to obtain detection window function $W(m_1, m_2; z)$. In principle, the window function that we obtain should be interpreted as an effective circularized-orbit approximation. Although PBH binaries were born with a broad eccentricity distribution, the present benchmark keeps only the circular orbit for the simplified SNR threhold, and we leave the eccentricity case for future work.

The next step is to calculate the redshift distribution of PBH binaries. Since we are using the comoving number density of PBH binaries that is assumed to be constant after their formation, the redshift distribution of PBH binaries equals the redshift distribution of comoving volume on the past light cone as,
\begin{align}\label{eq:redshift_dist_comoving_volume}
    p(z) =\frac{1}{V_c}  \frac{dV_c}{dz}~,
\end{align}
where $dV_c/dz$ is the differential comoving volume on the past light cone, and it reads as $dV_c/dz = 4 \pi D_H D_M^2/E(z)$ in the flat universe, where $D_H = c/H_0$ is Hubble distance, $D_M$ is transverse distance, and $E(z) \simeq \sqrt{\Omega_M (1+z)^3 + \Omega_\Lambda}$ in flat $\Lambda$CDM model \cite{Hogg:1999ad}. $\Omega_M$ and $\Omega_\Lambda$ is the energy density fraction of total matter and dark energy, and we use Planck 2018 data to set them as $(\Omega_M, \Omega_\Lambda) = (0.315, 0.685)$ in following calculation \cite{Planck:2018vyg}. $V_c$ is the total comoving volume on the past light cone, which can be calculated as $V_c = \int_0^\infty dV_c$.

\begin{figure}[htbp]
	\centering
	\includegraphics[width=10cm]{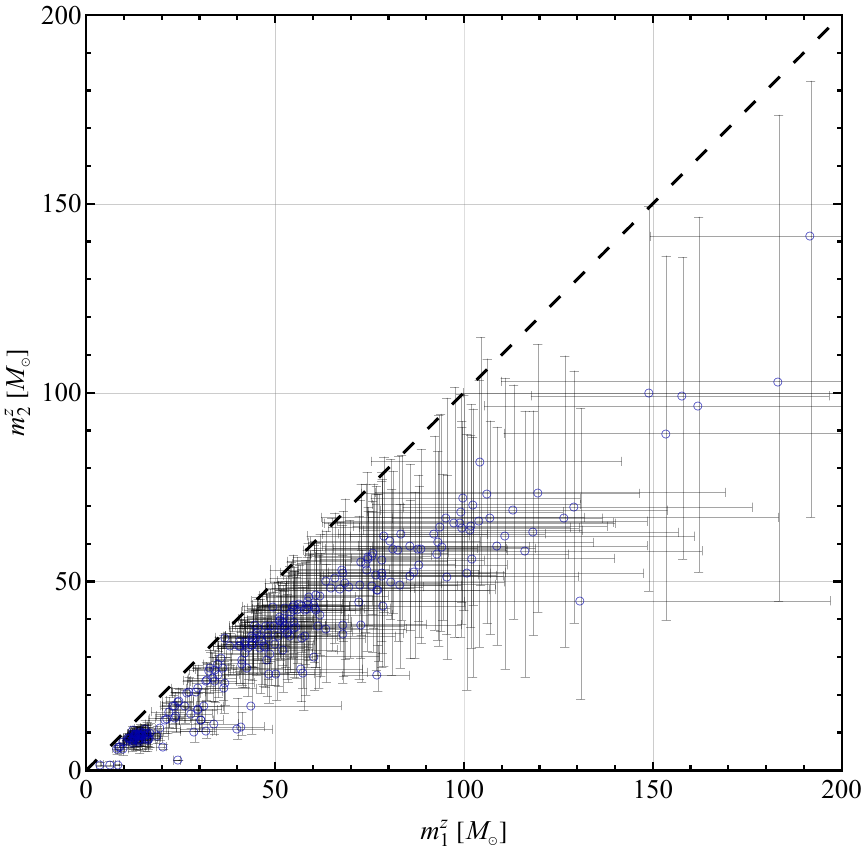}
	\caption{The redshifted mass of picked $256$ BBH events in LVK catalogs (GWTC-2.1-confident, GWTC-3-confident, GWTC-4.1, and GWTC-5.0). The selection criteria for BBH events are the value of signal-to-noise ratio ${\rm SNR} \geq 8$ and $p_{\rm astro} \geq 0.9$. Each BBH event has the median value of redshifted mass of its components $m_1^z$ and $m_2^z$ ($m_1^z > m_2^z$) with corresponding 90\% credible interval.}
	\label{fig:redshift_mass_dist}
\end{figure}

As the illustration of our method, we use LVK catalogs as a benchmark example. In LVK catalogs \cite{LIGOScientific:2018mvr, LIGOScientific:2020ibl, KAGRA:2021vkt, LIGOScientific:2025slb,  LIGOScientific:2026sit}, we pick up 256 events whose SNR is larger than a threshold value of $8$ and the probability of astrophysical events $p_{\rm astro} \geq 0.9$. Here $p_{\rm astro}$ describes the probability that the event is coming from the astrophysical phenomena rather than noise and it doesn't distinguish the stellar-origin and primordial-origin BHs. Each event has a corresponding redshifted mass pair $(m_{1}^z \pm_{\sigma_{1-}^z}^{\sigma_{1+}^z},m_{2}^z \pm_{\sigma_{2-}^z}^{\sigma_{2+}^z})$, where $m_i^z$ is the median and $\sigma_{i\pm}^z$ is the  90\% credible interval in its probability distribution. Their distribution on the $(m_{1}^{z},m_{2}^{z})$ plane is shown in Fig.~\ref{fig:redshift_mass_dist}. 
To model the probability distribution of the redshifted mass, we impose the split-normal distribution as an approximated distribution of masses involved each event, and the parameters of distribution are set as $\mu = m_i^z$ and $\sigma_{\pm} = \sigma_{i \pm}^z/1.645$, where we transfer 90\% credible interval $\sigma_{i \pm}^z$ in the redshifted mass to the standard deviation by dividing a factor $1.645$.
\begin{align}\label{eq:mass_probability}
    p(m_z) = 
    \left\{
    \begin{aligned}
		&\frac{\sqrt{2/\pi}}{\sigma_+ + \sigma_-} \exp\left(- \frac{(m_z - \mu)^2}{2 \sigma_+^2}\right)~,~m_z \geq \mu \\
		&\frac{\sqrt{2/\pi}}{\sigma_+ + \sigma_-} \exp\left(- \frac{(m_z - \mu)^2}{2 \sigma_-^2}\right)~,~m_z < \mu 
	\end{aligned}
    \right.
\end{align}
It should be noticed that by modeling $m_1^z$ and $m_2^z$ as independent split-normal distribution, we have neglected their covariance. Since the current LVK measurement uncertainty is large, this split-normal distribution can be viewed as a conservative choice to model the probability distribution of the redshifted mass.

However, not all of the LVK events are sourced from binary PBHs. To reconstruct the PBH mass function from detected BBH events, we need to filter out the potential PBH candidates. We introduce two simple selection criteria, based solely on mass, to remove potential astrophysical BBH sources just for illustration. The weaker selection criterion is to remove BHs with mass less than $15 \ M_\odot$ since in this mass range the population is usually dominated by Pop I and II BHs \cite{Tanikawa:2021qqi}. This selection criterion leaves us $174$ potential PBH binary events. The stronger selection criterion is to remove BHs with mass less than $50 \, M_\odot$, which is the lower bound of astrophysical BH mass gap in pair-instability mechanism \cite{2019ApJ...887...53F, Farmer:2020xne, Woosley:2021xba, Tong:2025wpz}, and this selection criterion leaves us $6$ potential PBH binary events in the LVK catalogs. Under both selection criteria, we assume the remaining events have a primordial origin and can be used in the following reconstruction.

\begin{figure}[htbp]
	\centering
	\includegraphics[width=8cm]{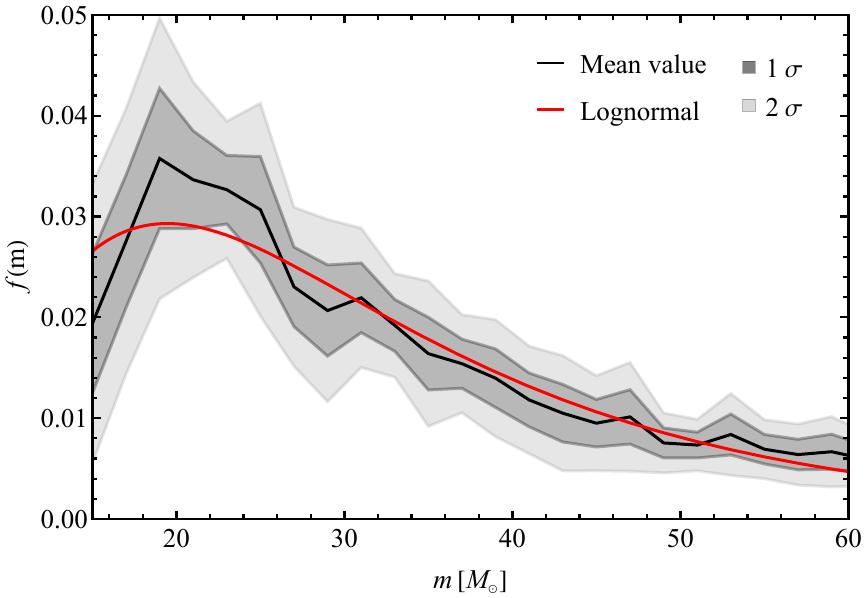}
    \includegraphics[width=8.2cm]{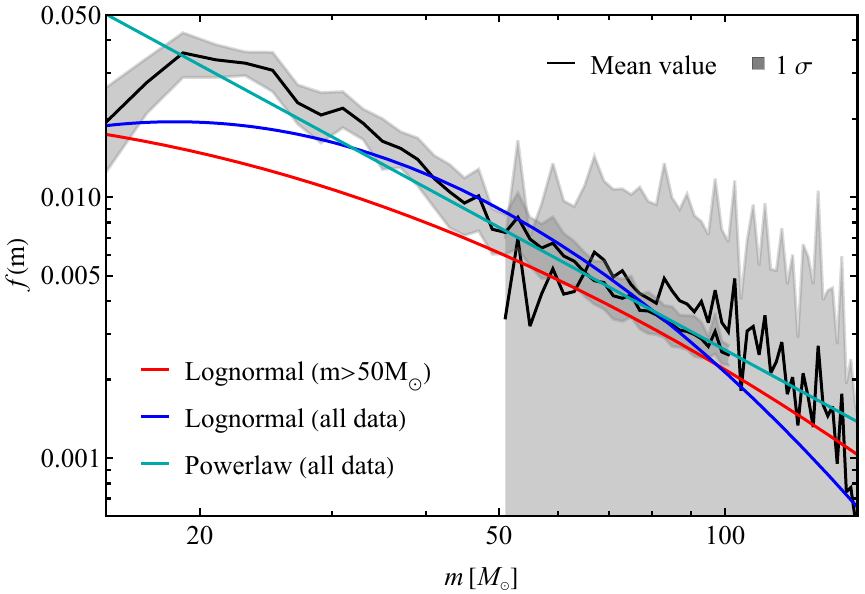}
	\caption{[Left]: The reconstructed PBH mass function from BBH events with an intrinsic mass larger than $15 \, M_\odot$ in LVK catalogs. The black curve with uncertainty region denotes the reconstructed result of PBH mass function. The red curve is best-fitting result of lognormal function with the mean value of the reconstruction.  [Right]: Two reconstructed PBH mass functions from BBH events with the intrinsic mass larger than $15 \, M_\odot$ and $50\,M_\odot$ in LVK catalogs. The red and blue curves are best-fitting result of lognormal function for the reconstructed PBH mass function in the mass range $(50\,M_\odot,150\,M_\odot)$ and $(15\,M_\odot,150\,M_\odot)$. The cyan curve is the best-fitting result of power-law function for the reconstructed PBH mass function in the mass range $(15\,M_\odot,150\,M_\odot)$. We should notice that selected BBH events are assumed to be PBH binaries for this demonstration.}
	\label{fig:PBH_reconstruct}
\end{figure}

Before using the above discussed method to reconstruct PBH mass function, we should first determine a PBH redshifted mass for each event based on Eq.~\eqref{eq:mass_probability}. This allows us to generate the redshifted mass pair $(m_1^z, m_2^z)$ for each PBH binary event, and this process produces various redshifted mass distribution for each round of PBH mass function reconstruction. Then we reconstruct PBH mass function from redshifted mass distribution via gradient-descent method. We repeat this process for $100$ times so that $100$ PBH mass functions will be produced. We set the number of round to be $100$, since it is large enough to allow us to statistically approach the true mass distribution of PBHs from LVK data. In this reconstruction process, we use BBH events with $m>15 \, M_\odot$ to reconstruct PBH mass function in the mass range $(1 \, M_\odot, \, 100 \, M_\odot)$, and use BBH events with $m > 50 \, M_\odot$ to reconstruct PBH mass function in the mass range $(50 \, \, M_\odot, 150 \, M_\odot)$. Based on these reconstructed PBH mass functions, we can obtain mean value and standard deviation at each mass point as shown in Fig.~\ref{fig:PBH_reconstruct}.

The left panel of Fig.~\ref{fig:PBH_reconstruct} demonstrates the reconstructed PBH mass function from $174$ BBH events that is based on PBH selection criteria $m > 15 \, M_\odot$. It shows a peak PBH population around mass of $20 \, M_\odot$. Meanwhile, the right panel demonstrates the two reconstructed PBH mass functions from both $174$ and $6$ BBH events in LVK catalogs, respectively. It clearly shows that the uncertainty in reconstructed mass function from $6$ BBH events is much larger than that from $174$ events because the sample contains fewer BBH events. Comparing the two reconstructions, their behaviors in the overlap mass range $(50 \, M_\odot, 100 \, M_\odot)$ are consistent with each other and this provides an internal consistency check of the benchmark reconstruction, under the assumption that the selected events are PBH candidates.

After reconstructing the PBH mass function from LVK data, we can use two kinds of PBH mass function to fit this reconstructed result. One is the lognormal function \cite{Dolgov:1992pu} and the other one is power-law function \cite{Green:1999xm} as
\begin{align}\label{eq:LN_PL}
    f_{\rm LN}(m)= \frac{1}{\sqrt{2 \pi} \sigma_{\mathrm{MF}} m} \exp\left(-\frac{\ln^2 (m/m_c)}{2 \sigma_{\mathrm{MF}}^2}\right)~,~~~f_{\rm PL}(m) = \frac{A_{\mathrm{MF}}}{M_\odot} \left(\frac{m}{M_\odot}\right)^{-\alpha_{\mathrm{MF}}}~.
\end{align}
Here, $m_c$ is the characteristic mass of PBHs and $\sigma_{\mathrm{MF}}$ is the width of PBH mass function. $\alpha_{\mathrm{MF}}$ is the power index of power-law PBH mass function, and $A_{\mathrm{MF}}$ is its dimensionless factor.
Then we can fit the reconstructed PBH mass function with these two types of mass function via minimizing the reduced $\chi^2$, which is defined as
\begin{align}\label{eq:reduced_chi_square}
    \chi_\nu^2 = \frac{1}{N-k}\sum_i^N \frac{(f_{\rm model}(m_i|\theta) - f_{\rm rec}(m_i))^2}{\sigma(m_i)^2}~,
\end{align}
where $\theta$ is the parameter in the fitting PBH mass function, $N$ is the number of data in fitting, and $k$ is the number of free parameters in the fitting function. Meanwhile, we can use 100 reconstructed PBH mass function as samples to fit the lognormal and power-law function, and this generates statistical values (mean value and standard deviation) of parameters for these two functions. The fitting results are demonstrated as solid curves in Fig.~\ref{fig:PBH_reconstruct} and the fitting parameters are summarized in Table.~\ref{tab:fitting_parameter}.

\begin{table}[htbp]
    \centering
    \caption{The fitting parameters of reconstructed BH mass function.}
    \renewcommand{\arraystretch}{1.2} 
    \begin{tabular}{c c | c c }
        \hline
        \hline
        \multicolumn{2}{c|}{Best lognormal fitting with selection criteria} & \multicolumn{2}{c}{Best fitting for all data} \\
        \hline
        Para ($m > 15 \, M_\odot$) & Value & Lognormal Para  & Value  \\
        \hline
        $m_c \, [M_\odot]$  & $27.5$               &  $m_c \, [M_\odot]$    & $35.2$           \\
        $\sigma_{\mathrm{MF}}$            & $0.59$               &  $\sigma_{\mathrm{MF}}$    & $0.80$            \\
        $\chi_\nu^2$            & $0.49$ &  $\chi_\nu^2$    & $0.75$         \\
        \cline{1-4}
        Para ($m > 50 \, M_\odot$)& Value       &  Power-law Para & Value   \\
        \cline{1-4}
        $m_c \, [M_\odot]$  & $33.5$              &  $\alpha_{\mathrm{MF}}$    & $1.56$ \\
        $\sigma_{\mathrm{MF}}$            & $1.24$               &  $A_{\mathrm{MF}}$    & $3.46$ \\
        $\chi_\nu^2$            & $0.07$   &  $\chi_\nu^2$ & $0.54$  \\
        \hline
        \hline
        \multicolumn{2}{c|}{~Statistical lognormal fitting with selection criteria~} & \multicolumn{2}{c}{~Statistical fitting for all data~} \\
        \hline
        Para ($m > 15 \, M_\odot$) & Value & Lognormal Para & Value  \\
        \hline
        $m_c \, [M_\odot]$    & $27.3 \pm 3.0 $    &  $m_c \, [M_\odot]$ & $33.5 \pm 2.8$               \\
        $\sigma_{\mathrm{MF}}$              & $0.54 \pm 0.04$    &  $\sigma_{\mathrm{MF}}$           & $0.73 \pm 0.10$               \\
        \cline{1-4}
        Para ($m > 50 \, M_\odot$)& Value        &  Power-law Para & Value          \\
        \cline{1-4}
        $m_c \,[M_\odot]$     & $108.5 \pm 76.3 $   &  $\alpha_{\mathrm{MF}}$ & $1.44 \pm 0.12$   \\
        $\sigma_{\mathrm{MF}}$              & $1.08\pm 0.38$     &  $A_{\mathrm{MF}}$ & $2.67 \pm 1.19$  \\
        \hline
        \hline
    \end{tabular}
    \label{tab:fitting_parameter}
\end{table}

In fitting reconstructed PBH mass function with selection criterion $m > 15 \, M_\odot$, we choose the PBH mass range $(15 \, M_\odot,\, 60 \, M_\odot)$ to fit the lognormal function as shown in the left panel of Fig.~\ref{fig:PBH_reconstruct}. Meanwhile, we use reconstructed PBH function with selection criterion $m > 50 \, M_\odot$ in mass range $(50 \, M_\odot, \, 150 \, M_\odot)$ to fit the lognormal function as shown in the right panel of Fig.~\ref{fig:PBH_reconstruct}. In the left panel, it clearly shows that the lognormal function fits the peak behavior in the reconstructed PBH mass function well, where lognormal function shows a characteristic PBH mass $m_c = 27.5 \, M_\odot$ and the width of the PBH mass function $\sigma_{\mathrm{MF}} = 0.59$. Meanwhile, lognormal fitting result in the right panel demonstrates similar trend in low mass range, which produces parameters $m_c = 33.5 \, M_\odot$ and $\sigma_{\mathrm{MF}} = 1.24$. Although their fitting parameters are not a perfect match, the consistent behavior of reconstructed PBH mass functions in the overlap mass range $(50 \, M_\odot, \, 100 \, M_\odot)$ still hints at the similarity of the fitting result, and this consistency is also shown in the similar trend of two reconstructed PBH mass functions in the large mass region. There are several issues that lead to unmatched fitting parameters, one is lognormal function may not be a suitable function to fit the global feature of reconstructed PBH mass function, the other one is the fewer samples in BBH events with $m > 50 \, M_\odot$ that causes a large error bar in reconstruction and leads to a much smaller value of $\chi_\nu^2$ than $1$. In order to check the consistency of fitting results with two selection criteria, we combine the reconstruction data from both selection criteria and use lognormal and power-law function to fit their global features as shown in the right panel of Fig.~\ref{fig:PBH_reconstruct}. The lognormal function fitting produces the similar result as those with selection criteria that is $m_c = 35.2 \, M_\odot$ and $\sigma_{\mathrm{MF}} = 0.80$, and the power-law function fitting produces a global feature on power index $\alpha_{\mathrm{MF}} = 1.56$.

The above discussion is to reconstruct normalized PBH mass function $f(m)$ from LVK catalogs, then we can use the reconstructed PBH mass function to calculate energy density fraction of PBHs in dark matter $f_{\rm PBH}$ based on the number of PBH binary mergers. The volumetric merger rate density of early PBH binaries can be estimated as \cite{Raidal:2018bbj}
\begin{align} \label{eq:merger_rate_density}
    \frac{R(z)}{dm_1 dm_2} = \frac{1.6 \times 10^6}{{\rm Gpc^3 \, yr}} f_{\rm PBH}^{\frac{53}{37}} \left(\frac{m_1 m_2}{(m_1+m_2)^2}\right)^{-\frac{34}{37}} \left(\frac{m_1+m_2}{M_\odot}\right)^{-\frac{32}{37}} \left(\frac{t(z)}{t_0}\right)^{-\frac{34}{37}} \frac{m_1 m_2}{\langle m \rangle^2} f(m_1) f(m_2) S(f_{\rm PBH})~,
\end{align}
where $S(f_{\rm PBH})$ is the suppression factor that can be approximated as $S = (1 + \sigma_M^2/f_{\rm PBH}^2)^{-21/74}$ with $\sigma_M = 8.5 \times 10^{-2}$ is the rescaled variance of matter density perturbations at the PBH binary formation time \cite{Raidal:2018bbj, Ali-Haimoud:2017rtz}. $\langle m \rangle$ is the average mass of PBH, and it can be calculated as
\begin{align}
    \langle m \rangle = \frac{\rho_{\rm PBH}}{n_{\rm PBH}} = \frac{\int m dn}{n_{\rm PBH}} = \int m \frac{1}{n_{\rm PBH}} \frac{dn}{dm} dm = \int m f(m) dm~. 
\end{align}
It gives a result of $\langle m \rangle = 36.7\,M_\odot$ using the mean value of reconstructed PBH mass function $f(m)$ based on $174$ events in LVK catalogs.
Then we integrate Eq.~\eqref{eq:merger_rate_density} over comoving volume $V_c$ and run duration time $T_{\rm obs}$ to obtain the merger events of PBH binaries. Consider the detection of PBH binaries in LVK, we impose the SNR larger than 8 to obtain the number of detectable PBH merger events $N_{\rm PBH}$ as
\begin{align}\label{eq:observed_pbh_num}
    N_{\rm PBH} = \int_{z}\int_{m_1}\int_{m_2} \frac{R(z)}{dm_1 dm_2} \frac{T_{\rm obs}}{1+z}\frac{dV_c}{dz} \Theta({\rm SNR - 8}) dm_1 dm_2 dz~.
\end{align}
The integration range of redshift is set as $(0,1)$ that is the detection redshift range in LVK catalogs. $T_{\rm obs}$ is run duration of LVK in O1--O4, which is $2.11\,{\rm yrs}$ \cite{LIGOScientific:2018mvr, LIGOScientific:2020ibl, KAGRA:2021vkt, LIGOScientific:2025slb,  LIGOScientific:2026sit}, and we divide it by the redshift factor $1+z$ to calculate duration in the rest frame of PBH binaries. To determine the value of $f_{\rm PBH}$, we treat $174$ BBH events based on PBH selection criterion $m > 15 \, M_\odot$ as PBH candidates, and solve equation $N_{\rm PBH} (f_{\rm PBH}) = 174$. It gives the result as $f_{\rm PBH} = 1.08 \times10^{-3}$. 

In the evaluation of $f_{\rm PBH}$, we use the volumetric merger rate density in Eq.~\eqref{eq:merger_rate_density} rather than using our formulism in Eq.~\eqref{eq:probability}. Since the detected PBH binaries in LVK catalogs are all merged events, using volumetric merger rate density remains highly robust and computationally optimized for determining $f_{\mathrm{PBH}}$. Meanwhile, it should be emphasized that our method is targeting on high redshift PBH binaries where not only merged events but also inspiral events can be detected in future GW observations, hence we don't use Eq.~\eqref{eq:merger_rate_density} that only includes merged events in our method.

\section{Reconstruction of the Primordial Power Spectrum} \label{sec:PPS_reconstruction}

We first revisit the Press-Schechter method to compute the abundance of the PBH from primordial power spectrum (PPS) \cite{Green:2004wb}. Assuming that the PBH forms when the density perturbation within the horizon exceeds the threshold $\delta_\mathrm{c}$\footnote{$\delta_\mathrm{c}=0.45$ is used in the reconstruction.}, and the probability distribution of the density perturbation follows Gaussian distribution.

First, we smooth the density perturbation $\delta$ on a certain scale $R$ with a window function $W(R;x-x')$
\begin{equation}\label{eq:delta_R}
    \delta_{R}(x)=\int {d^3 x'}\delta (x')W(R;x-x').
\end{equation}
The variance of $\delta$ after smoothing is thus given
\begin{equation}\label{eq:sigma_square}
    \sigma^2(R)=\int_{0}^{\infty}\frac{dk}{k}\mathcal{P}_{\delta}(k)\tilde{W}^{2}(R;k),
\end{equation}
where $\mathcal{P}_{\delta}(k)$ is the power spectrum of the density perturbation, which can be derived from the power spectrum of comoving curvature perturbation generated from inflation , i.e. primordial power spectrum (PPS) by:
\begin{equation}\label{eq:P_delta}
    \mathcal{P}_{\delta}(k)=\frac{4(1+w)^2}{(5+3w)^2}(kR)^4\mathcal{P}_{\mathcal{R}}(k),
\end{equation}
where $w$ is the equation of state of the epoch during which PBHs form. 
Throughout the paper, we use a volume-normalized Gaussian window function, appears as
\begin{equation}
    \tilde{W}(R,k)=\exp\left(-\frac{k^2R^2}{2}\right)
\end{equation}
in the Fourier space. Therefore, we could summarize this step as \textit{one formula}
\begin{align}
    &\sigma^2(R)=\int_{0}^{\infty}{d\ln k}K(k,R)\mathcal{P}_{\mathcal{R}}(k),\\
    &K(k,R)=\frac{4(1+w)^2}{(5+3w)^2}(kR)^4\exp\left(-k^2R^2\right).
    \label{eq:pras1eq}
\end{align}
$\sigma$ here is a function of $R$, which determines the mass of the PBH, following
\begin{align}\nonumber\label{eq:m_R}
        m(R)&=\gamma_m M_{H}=\frac{4\pi}{3}\gamma_m\rho H^{-3}\\
        &\simeq 30M_{\odot}\left(\frac{\gamma_m}{0.2}\right)\left(\frac{g_{\star}}{10.75}\right)^{-1/6}\left(\frac{R}{3.3\times 10^{-6}\mathrm{Mpc}}\right)^{2}.
\end{align}
where the comoving scale $R$ is inversely proportional to the comoving wavenumber $k$, $g_{\star}$ is the number of relativistic degrees of freedom, and $\gamma_m$ is PBH collapse efficiency factor. For a fixed mass of PBH, the kernel $K(k,R)$ filters out the contribution from large $k$ modes.
To compute the PBH abundance from the $ \sigma^2(R)$, assuming $\delta$ follows Gaussian distribution (see discussions on perturbative non-Gaussianity in Sec.~\ref{nongaussianity}), we integrate the probability distribution function of $\delta$ for $\delta_{R}>\delta_\mathrm{c}$, 
\begin{equation}
\label{betasigma}
    \beta(R)=\int_{\delta_c}^{\infty}\frac{1}{\sqrt{2\pi}\sigma(R)}\exp\left(-\frac{\delta_{R}^2}{2\sigma^2(R)}\right)d\delta_R=\frac{1}{2}\mathrm{erfc}\left(\frac{\delta_c}{\sqrt{2}\sigma (R)}\right).
\end{equation}
We approximate the PBH mass function by identifying the collapse fraction $\beta(M)$ with the abundance in each mass bin, which is valid for sharply peaked spectra.
A more accurate treatment requires extracting the differential mass function via the derivative of the collapse fraction or using the excursion-set formalism, we leave it for future work.

According to the forward computation, we split the inverse problem to the following 3 steps.

\subsection{(i) From $f(m)$ to $\beta(R)$}
We prepare the data by converting $f_{\mathrm{PBH}}(m)$ into $\beta(m)$.
From the previous steps, the normalized PBH mass function is obtained $f(m)= n_\mathrm{PBH}^{-1} \frac{dn}{dm}$. However, the reconstruction will involve the mass function that indicates also the PBH fraction in dark matter $f_{\mathrm{PBH}}(m) \equiv df_{\rm PBH}/d\ln m=f_{\mathrm{PBH}}m^2f(m)/\langle m\rangle$, where $\langle m\rangle\equiv\int mf(m)dm$ is the averaged PBH mass, therefore we can obtain
\begin{align}\label{eq:fpbh_m_normalization}
    \int f_{\mathrm{PBH}}(m)m^{-1} ~{d m}\equiv f_{\mathrm{PBH}}\approx 1.08\times 10^{-3},
\end{align}
as derived in the previous section. 

We thus can obtain the mass function through \cite{Sasaki:2018dmp, Tomberg:2024chk}:
\begin{align}\label{eq:fpbh_m}
         f_{\mathrm{PBH}}(m)\simeq\left(\frac{\beta(m)}{3.7\times10^{-9}}\right)\left(\frac{10.75}{g_\star(T_{\mathrm{form}})}\right)^{1/4}\left(\frac{0.264}{\Omega_{\mathrm{DM}}}\right)\left(\frac{M_{\odot}}{m}\right)^{1/2}\left(\frac{\gamma_m}{0.2}\right)^{1/2},
\end{align}
where $T_{\mathrm{form}}$ is the temperature at PBH formation.

\subsection{(ii) From $\beta(R)$ to $\sigma^2(R)$}
This process also only involves basic arithmetic, following Eq.~\eqref{betasigma}, we obtain
\begin{equation}
    \sigma^2(m)=\frac{\delta_{\mathrm{c}}^2}{2[\mathrm{erfc}^{-1}(2\beta(m))]^2}.
\end{equation}
Once $\sigma^2(m)$ is obtained, we proceed to the key step:

\subsection{(iii) From $\sigma^2(R)$ to $\mathcal{P}_{\mathcal{R}}(k)$}
This step is the core part of the reconstruction, rather ill-posed. 
According to Eq.~\eqref{eq:pras1eq}, the problem is an inverse convolution which is not stable, since the inverse could amplify the small, high-frequency noises in the reconstructed PPS.
To solve $\mathcal{P}_\mathcal{R}$ , we first discretize the integral. The discretized kernel is defined as
\begin{align}
    K_{ij}=K(k_j,R_i)\Delta(\ln k_j),
\end{align}
where $N_R$ points in $R$ space and $N_{k}$ points in $k$ space are taken.
We can thus rewrite the integral as a sum
\begin{align}
\sigma_{i}^2=\lim_{N_{k}\rightarrow\infty}\sum_{j=1}^{N_{k}}K_{ij}\mathcal{P}_{j},
\end{align}
or in the matrix form
\begin{align}\label{eq:vector_sigma_square}
    \vec{\sigma^2}=\mathbf{K}\vec{\mathcal{P}}.
\end{align}
where $\mathcal{P}$ is the discretized PPS that we are targeting at.
The above equation has the solution given
\begin{equation}
   \vec{\mathcal{P}}=\mathbf{K}^{-1} \vec{\sigma^2},
\end{equation}
which is not stable if $\mathbf{K^{-1}}$ has a huge norm that amplifies the error. 
One way to resolve the problem is to apply the Tikhonov regularization. In short, the regularization is introduced to suppress unphysical oscillatory solutions and to make the inverse problem well-posed by providing additional prior information beyond the data. It allows the kernel to have a bounded inverse.
In matrix terms, the regularization lifts the near-null eigenmodes of the kernel-induced normal matrix, stabilizing the inversion and ensuring invertibility.

The reconstruction is therefore formulated as a minimization problem,
\begin{align}
    \vec{\mathcal P}_\lambda
    =
    \arg\min_{\vec{\mathcal P}}
    \left[
    \|\mathbf{K}\vec{\mathcal P}-\vec{\sigma^2}\|_2^2
    +
    \lambda \|\mathbf{L}\vec{\mathcal P}\|_2^2
    \right],
    \label{eq:minimization}
\end{align}
where $||X||_2=\sqrt{X^{T}X}$ denotes the Euclidean norm of $X$, $\mathbf{L}$ is determined by the regularization assumption introduced later, and $\lambda$ is a hyperparameter to control the strength of the smoothness prior. It controls the relative weight between the data misfit term and the smoothness penalty in the Tikhonov functional. We state the principle of the choice of $\lambda$ in Appendix~\ref{app:regularization} consider L-curve method. 

\begin{figure}[htbp]
    \includegraphics[width=.45\textwidth]{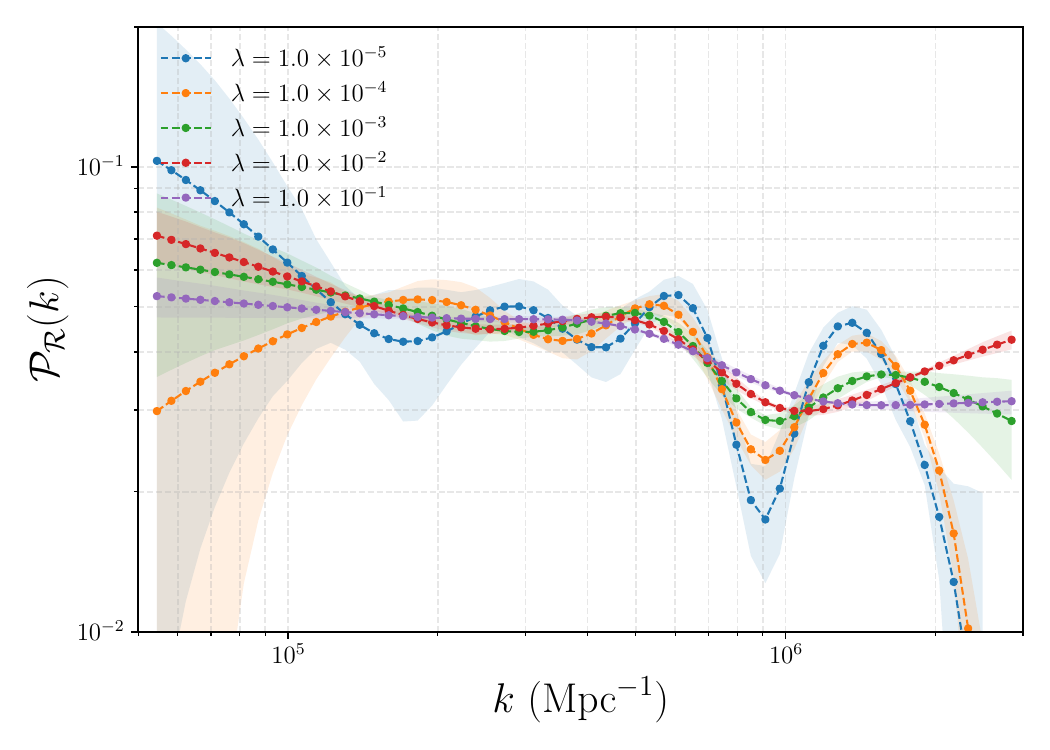}
    \includegraphics[width=.45\textwidth]{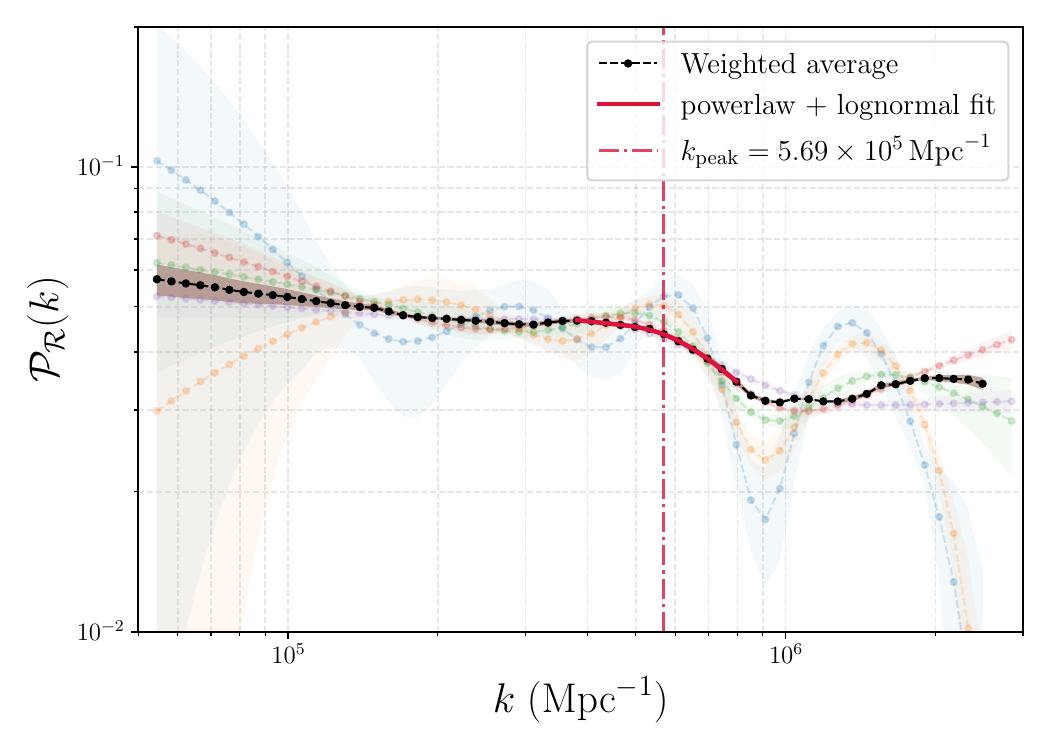}
    \includegraphics[width=.45\textwidth]{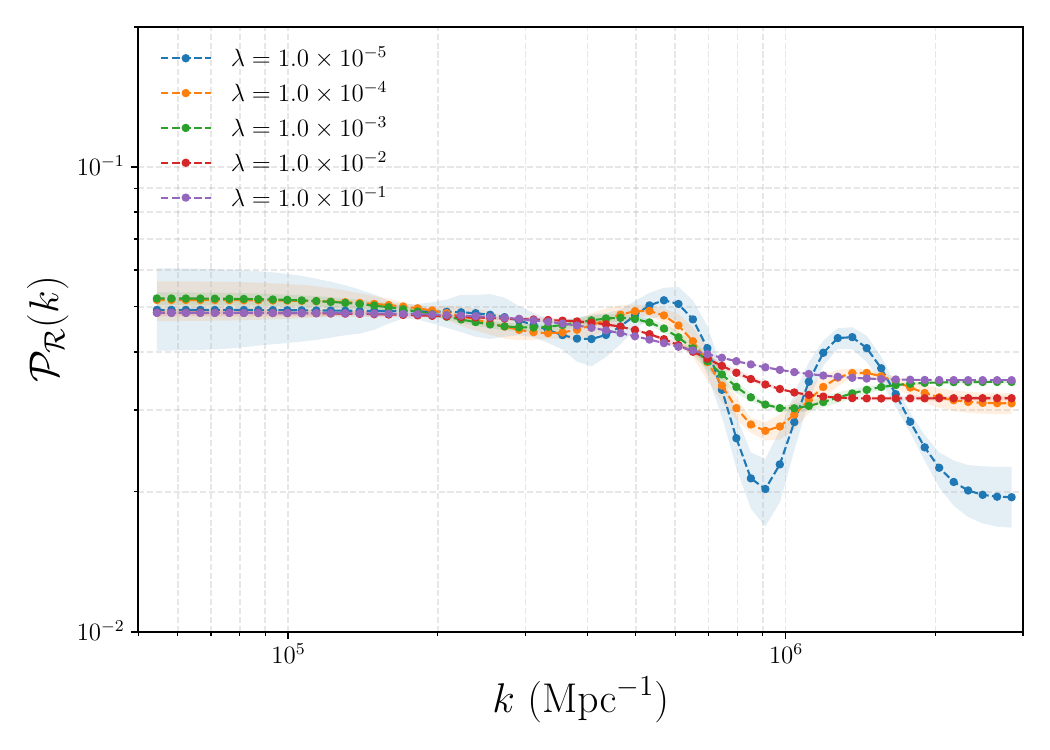}
    \includegraphics[width=.45\textwidth]{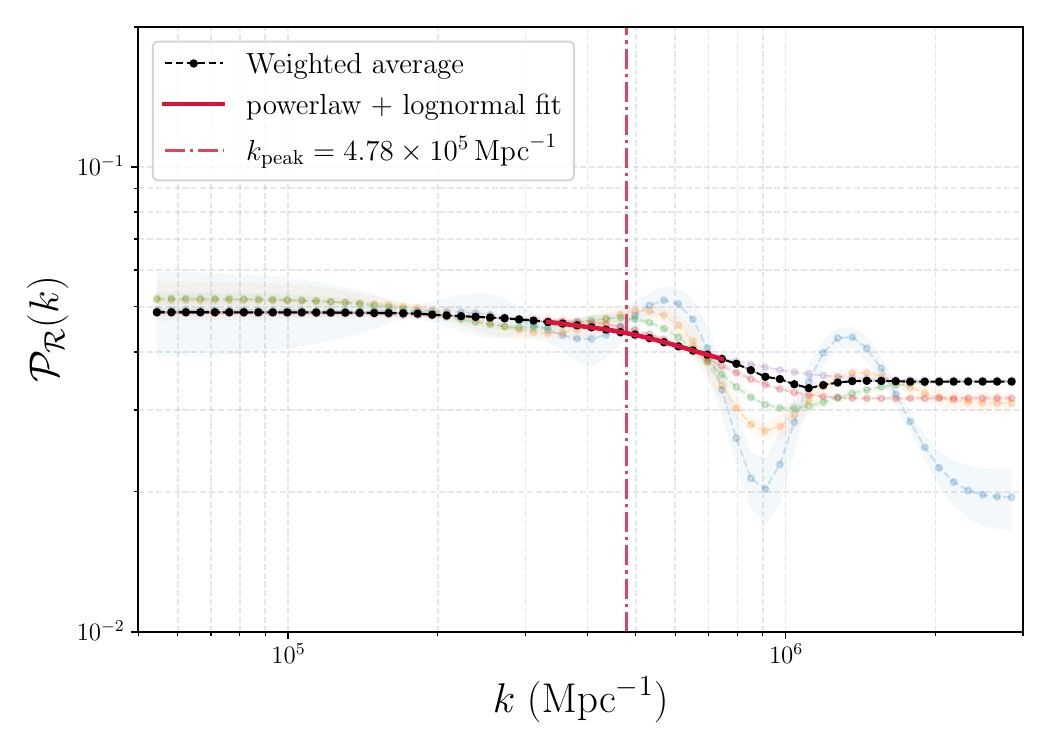}
	\caption{Reconstructed primordial power spectra $\mathcal{P}_{\mathcal{R}}(k)$ for different values of the regularization parameter $\lambda$ and assumptions (Upper panels: 2nd order, lower panels: 1st order). The left panels show the results for $\lambda=10^{-5},10^{-4},10^{-3}, 10^{-2},10^{-1}$ with $1\sigma$ error bar. The right panels show the weighted-averaged result (black dots) for the previous choices of $
    \lambda$ and the best-fit power-law plus lognormal model (red solid line). The red dashed line marks the best-fit template peak position for lognormal parameterization.}
	\label{fig:PPS_reconstruct}
\end{figure}

In practice, we make use of the following assumption of regularization
\begin{itemize}
    \item There is no large oscillation: $d^2\mathcal{P}_{\mathcal{R}}/d(\ln k)^2$ is small.
\end{itemize}
Accordingly, we implement the second-order regularization operator:
\begin{equation}
\mathbf{L}\mathcal{P}_i=\mathcal{P}_{i}-2\mathcal{P}_{i+1}+\mathcal{P}_{i+2}.
\end{equation}
In comparison, we also apply another assumption
\begin{itemize}
    \item There is no large local tilt: $d\mathcal{P}_{\mathcal{R}}/d(\ln k)$ is small.
\end{itemize}
Accordingly, we also implement the first-order regularization operator:
\begin{equation}
\mathbf{L}\mathcal{P}_i=\mathcal{P}_{i+1}-\mathcal{P}_{i}.
\end{equation}
Notably, taking into account the regularization introduces a mild model dependence to the inverse problem. As a result, the minimization problem Eq.~\eqref{eq:minimization} leads to a regularized normal equation
\begin{align}
\frac{\partial}{\partial\vec{\mathcal{P_\lambda}}}\left[(\mathbf{K}\vec{\mathcal{P_\lambda}}-\vec{\sigma^2})^{T}(\mathbf{K}\vec{\mathcal{P_\lambda}}-\vec{\sigma^2})+\lambda(\mathbf{L}\vec{\mathcal{P_\lambda}})^{T}(\mathbf{L}\vec{\mathcal{P_\lambda}})\right]=0,
\end{align}
whose solution is
\begin{align}\label{eq:solution_P_lambda}
    \vec{\mathcal{P_\lambda}}=\mathbf{R}_{\lambda}\vec{\sigma^2},
\end{align}
where the reconstruction matrix is defined as $\mathbf{R}\equiv(\mathbf{K}^{T}\mathbf{K}+\lambda \mathbf{L}^{T}\mathbf{L})^{-1}\mathbf{K}^{T}$, which is bounded due to the regularization. For each mass function $f(m)$ reconstructed in the previous step, we perform the PPS reconstruction using several choices of the regularization parameter $\lambda$. The results are shown in the left panels of Fig.~\ref{fig:PPS_reconstruct}. Here, we estimate the uncertainty band directly from the distribution of reconstructed spectra, by taking the 16th and 84th percentiles at each $k$, i.e.
\begin{align}\label{eq:sigma_lambda}
\sigma_{\lambda,j}^{-}=\mathcal{P}_{\lambda,j}^{50\%}-\mathcal{P}_{\lambda,j}^{16\%}\\
\sigma_{\lambda,j}^{+}=\mathcal{P}_{\lambda,j}^{84\%}-\mathcal{P}_{\lambda,j}^{50\%}.
\end{align}

We combine the reconstructed spectra obtained with different $\lambda$ using inverse-variance weighting with the uncertainty of weighted mean ideally estimated as
\begin{align}\sigma_{\langle\mathcal{{P}}\rangle}(k_{j})=\left(\sum_{\lambda}\sigma^{-2}_{\lambda,j}\right)^{-1/2},
\end{align}
where 
\begin{equation}
\sigma_{\lambda,j}\equiv\frac{1}{2}\left(\sigma_{\lambda,j}^{-}+\sigma_{\lambda,j}^{+}\right)=\frac{\mathcal{P}_{\lambda,j}^{84\%}-\mathcal{P}_{\lambda,j}^{16\%}}{2}.
\end{equation}

However, since the results are derived from the same underlying dataset, they are generally not statistically independent. Therefore, the uncertainty should be regarded as a phenomenological measure of the scatter across $\lambda$. A more strict statistical $1\sigma$ confidence interval will largely depend on the smallest $\lambda$ allowed, i.e. the interval should be no smaller than the largest one for an individual choice of $\lambda$. 
We use a power-law plus lognormal-shaped peak to parameterize the weighted-average result, 
\begin{align}\label{eq:fitting_fun_shape}
    \mathcal{P}_{\mathcal{R}}(k)=\alpha_{p}\left(\frac{k}{k_{\mathrm{peak}}}\right)^{n_p}+\beta_p\exp\left[-\frac{1}{2}\left(\frac{\log_{10}(k/k_{\mathrm{peak}})}{\sigma_p}\right)^2\right].
\end{align}
The fitting results are demonstrated in the right panels of Fig.~\ref{fig:PPS_reconstruct} and in Table.~\ref{tab:fitting_parameter_pr}. 
The apparent non-decaying behavior at the IR boundary should not be interpreted as a physical prediction of the PPS at lower wavenumbers. The reconstruction is only constrained within the $k$-range mapped from the reconstructed PBH mass interval. Since smaller $k$ corresponds to larger PBH masses, an extrapolation of the IR tail beyond this range would in general produce additional massive PBHs and could overestimate the total PBH abundance.

\begin{table}[htbp]
\centering
\begin{tabular}{c|c|c}
\hline
Parameter & 2nd-order regularization best-fit & 1st-order regularization best-fit \\
\hline
$\alpha_p$ & $0.035 \pm 0.007$ & $0.042 \pm 0.001$ \\
$n_p$ & $-0.50 \pm 0.17$ & $-0.22 \pm 0.02$ \\
$\beta_p$ & $0.0082 \pm 0.0064$ & $0.0020 \pm 0.0011$ \\
$k_{\rm peak}/{10^5\rm Mpc}^{-1}$ & $5.69^{+0.33}_{-0.31}$  & $4.78^{+0.25}_{-0.23}$ \\
$\sigma_p$ & $0.13 \pm0.06$ & $0.13\pm 0.05$ \\
$\chi^2_\nu$ & $0.17$ & $0.17$ \\
\hline
\end{tabular}
\caption{Fitting parameters of reconstructed PPS with optimistic $1\sigma$ confidence intervals for the weighted-averaged result shown in the right panels of Fig.~\ref{fig:PPS_reconstruct}.}
\label{tab:fitting_parameter_pr}
\end{table}

\subsection{Non-Gaussianity}
\label{nongaussianity}
So far, the reconstruction has been performed under the assumption that the smoothed density contrast follows a Gaussian distribution. This assumption provides a minimal and commonly used baseline, but the abundance of PBHs is exponentially sensitive to the tail of the probability distribution. Even a small primordial non-Gaussianity can therefore modify the mapping between the collapse fraction \(\beta(M)\) and the variance \(\sigma^2(R)\), and consequently affect the reconstructed primordial power spectrum. 

\begin{figure}[htbp]
    \includegraphics[width=.45\textwidth]{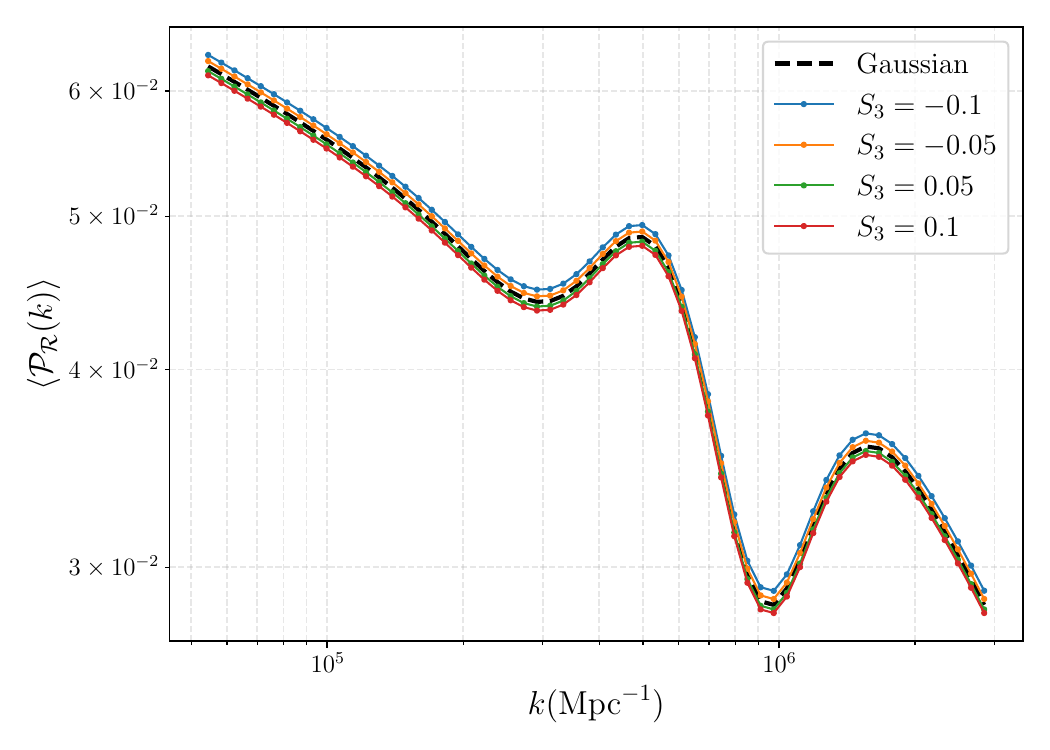}
     \includegraphics[width=.45\textwidth]{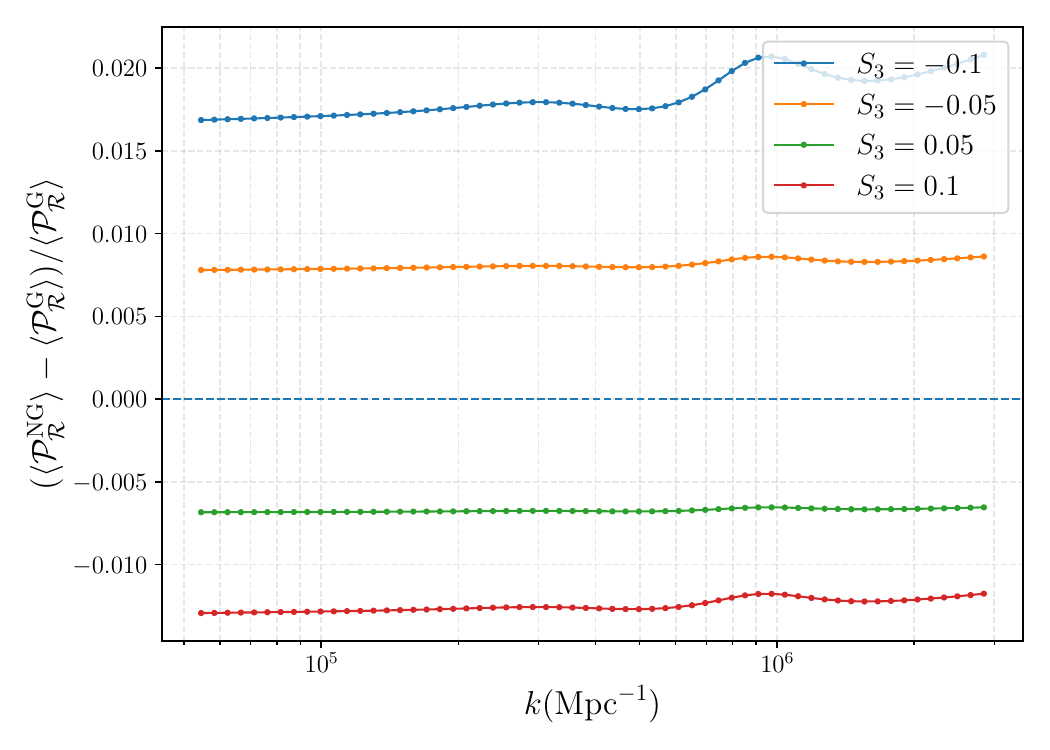}
	\caption{Mean value of reconstructed PPS and differences between Gaussian and non-Gaussian reconstructions for different values of $S_3$. Here, second-order regularization with $\lambda=10^{-3}$ is taken.}
	\label{fig:PPS_reconstruct_ng}
\end{figure}

In this subsection, we examine how weak non-Gaussian corrections can be incorporated perturbatively. Rather than committing to a specific model of non-Gaussianity, we parameterize the one-point probability distribution of the smoothed density contrast by an Edgeworth expansion
\begin{align} \label{eq:Edgeworth expansion}
    P(\delta_{R})=P_G(\delta_R)\left(1+\frac{\sigma S_3}{6}H_3\left(\frac{\delta_R}{\sigma}\right)+\frac{\sigma^2S_4}{24}H_4\left(\frac{\delta_R}{\sigma}\right)+\frac{\sigma^2S_3^2}{72}H_{6}\left(\frac{\delta_R}{\sigma}\right)+\cdots\right),
\end{align}
where
\begin{align}
    P_G(\delta_R)\equiv\frac{1}{\sqrt{2\pi}\sigma(R)}\exp\left(-\frac{\delta_{R}^2}{2\sigma^2(R)}\right),
\end{align}
 and $H_{n}$ are the probabilists’ Hermite polynomials, given by
\begin{align}
    &H_{n}(\nu)\equiv(-1)^ne^{\nu^2/2}\frac{d^n}{d\nu^n}e^{-\nu^2/2},\\
    &H_3(\nu)=\nu^3-3\nu,\\
    &H_4(\nu)=\nu^4-6\nu^2+3,\\
    &H_6(\nu)=\nu^6-15\nu^4+45\nu^2-15.
\end{align}
Here $S_3$ and $S_4$ are the skewness and kurtosis, defined by
\begin{align}
    &S_3\equiv\frac{\langle \delta_R^3 \rangle_c}{\sigma^4},\\
    &S_4\equiv\frac{\langle \delta_R^4 \rangle_c}{\sigma^6}.
\end{align}
As a result, using the property of Hermite polynomials:
\begin{align}
\int_{\delta_c}^{\infty}P_{G}(\delta_R)H_{n}\left(\frac{\delta_R}{\sigma}\right)d{\delta_R}=\sigma P_{G}(\delta_c)H_{n-1}\left(\frac{\delta_c}{\sigma}\right),
\end{align}
then abundance Eq.~\eqref{betasigma} obtained from forward computation changes into 
\begin{align}\nonumber
    \beta(R)&=  \int_{\delta_c}^{\infty}P_G(\delta_R)\left(1+\frac{\sigma S_3}{6}H_3\left(\frac{\delta_R}{\sigma}\right)+\frac{\sigma^2S_4}{24}H_4\left(\frac{\delta_R}{\sigma}\right)+\cdots\right)d\delta_R\\
    &=\frac{1}{2}\mathrm{erfc}\left(\frac{\delta_c}{\sqrt{2}\sigma (R)}\right)+P_{G}(\delta_c)\left(\frac{\sigma^2 S_3}{6}H_2\left(\frac{\delta_c}{\sigma}\right)+\frac{\sigma^3S_4}{24}H_3\left(\frac{\delta_c}{\sigma}\right)+\frac{\sigma^3S_3^2}{72}H_5\left(\frac{\delta_c}{\sigma}\right)+\cdots\right).
\end{align}
In the reconstruction We retain only the leading skewness correction, which indicates
\begin{align}
    \beta(R)=\frac{1}{2}\mathrm{erfc}\left(\frac{\delta_c}{\sqrt{2}\sigma }\right)+P_{G}(\delta_c)\frac{\sigma^2 S_3}{6}H_2\left(\frac{\delta_c}{\sigma}\right).
\end{align}
However, the inclusion of non-Gaussianity can make the reconstruction nontrivial. In the Gaussian case, the collapse fraction $\beta(R)$ is a monotonic function of the standard deviation $\sigma(R)$, so that it can be uniquely determined from a given $\beta(R)$. Once non-Gaussian corrections are included, this monotonicity is not automatically guaranteed. In particular, the perturbative correction to the tail probability may cause the relation between $\beta$ and $\sigma$ to become non-monotonic, leading to a multi-valued reconstruction of $\sigma(R)$. 
So in order to avoid the ambiguity, we impose the condition $|\sigma S_3H_3\left(\frac{\delta_R}{\sigma}\right)|/6\ll 1,~\partial\beta/\partial\sigma>0,$ to $S_3$. Since our data satisfies $\sigma<\sqrt{-1+\sqrt{2}}\delta_c$, the conditions indicate 
\begin{align}
 S_3 &>\max\left(\frac{6\delta_c\sigma^2}{-\delta_c^4+2\delta_c^2\sigma^2+\sigma^4}\right)\\
 |S_3|&\ll\min\left( \frac{6}{\sigma H_3\left(\frac{\delta_R}{\sigma}\right)}\right),\, 
\end{align}
which guarantees both that 1) the first non-Gaussian correction is perturbative and 2) the reconstruction is single-valued in the range of the reconstructed $\sigma$. We perform the sanity check after every reconstruction. Considering a scale-independent skewness, we obtained the reconstruction result for different values of $S_3$ for a fixed regularization strategy (see in Fig.~\ref{fig:PPS_reconstruct_ng}). 
The trend shown in Fig.~\ref{fig:PPS_reconstruct_ng} has a simple physical interpretation. 
Since a positive skewness enhances the high-density tail, the same collapse fraction $\beta(M)$ in this case should be obtained with a smaller variance $\sigma^2(R)$ than in the pure Gaussian case. As $\sigma^2(R)$ is sourced by the primordial curvature power spectrum through Eq.~\eqref{eq:pras1eq}, the reconstructed $\mathcal P_{\mathcal R}(k)$ is slightly reduced for positive $S_3$. Conversely, a negative skewness suppresses the high-density tail, requiring a larger variance and hence a slightly enhanced reconstructed power spectrum. 

\section{Conclusions} \label{sec:conclusion}
In this work, we propose a new framework to reconstruct the primordial black hole (PBH) mass function and subsequently the small-scale primordial power spectrum (PPS), from gravitational wave observations of binary black holes (BBHs) with redshift information. Starting from the observed redshifted mass distribution, we derived its connection to the underlying PBH mass function through the detector window function and the redshift distribution of PBH binaries. We then showed that this inverse problem can be solved numerically with a gradient-descent method, and that the reconstructed mass function can be mapped to the PBH abundance and the PPS by combining the Press-Schechter formalism with Tikhonov regularization. Our analysis, therefore, demonstrates that gravitational-wave observations of high-redshift BBHs can provide a new probe of primordial fluctuations on scales far smaller than those accessible to the cosmic microwave background or large-scale structure observations.

As a proof of concept, we applied this method to the LVK catalogs with simplified mass-selection cuts. Using two selection criteria, $m> 15 \,M_{\odot}$ and $m>50 \, M_{\odot}$, to remove possible stellar-origin BHs, we reconstructed the candidate PBH mass function from 174 and 6 events for each criterion, respectively. The reconstruction based on the $m> 15 \, M_{\odot}$ sample is well-modeled by a lognormal function, peaking at a characteristic mass $27.5 \, M_{\odot}$, while reconstruction based on the $m > 50 \, M_\odot$ sample demonstrates similar behavior, despite the much larger uncertainty in this mass interval due to limited statistics. Two independently reconstructed mass functions remain broadly consistent in the overlap region $50\sim 100 \, M_{\odot}$ and shows a consistent power index around $1.56$ in the whole PBH mass region $15 \sim 150 \, M_\odot$. Matching the observed merger abundance yields an estimated total PBH fraction in dark matter $f_{\mathrm{PBH}}\simeq 1.08\times 10^{-3}$ within the mass range of our concern, which is allowed under the current PBH constraint from LVK catalogs \cite{Andres-Carcasona:2024wqk, Bouhaddouti:2025ltb, Andres-Carcasona:2026avd}. It should be noted that this LVK application uses the simplified PBH candidate selection, therefore, these numerical results should be interpreted as an illustration of the benchmark model rather than as a definitive PBH population.

Based on the reconstructed PBH population, we obtained a slightly model-dependent reconstruction of the PPS. The result shows a candidate bump-like enhancement of order $\mathcal{O}(10^{-2})$ around $k_{\mathrm{peak}}\simeq 5.7\times 10^5~\mathrm{Mpc}^{-1}$ under the adopted assumptions. Its stability across the regularization choices considered suggests that the feature is connected to the peak-like structure in the reconstructed PBH mass function.
We also examined weak non-Gaussian corrections and found that the reconstruction procedure can be consistently extended beyond the pure Gaussian case.

Looking ahead, the reconstruction strategy proposed in this work is particularly well suited for next-generation gravitational-wave detectors such as Einstein Telescope, DECIGO, LISA, and TianQin/Taiji. High-redshift BBH events, especially those at $z\gtrsim 20$, would strongly reduce the contamination from conventional astrophysical BHs and provide a cleaner sample for probing PBH populations. With sufficient statistics, the redshifted mass distribution of such events could therefore be used not only to reconstruct the PBH mass function, but also to infer the small-scale primordial curvature spectrum in a data-driven way.

\section*{Acknowledgments}
X.W. thanks Elisa Ferreira, Masahiro Takada, Sachiko Kuroyanagi, Alexander Kusenko and Misao Sasaki for useful discussions.
This work is supported by IBS under the project code, IBS-R018-D3 [Q.D. and M.Y.], by Forefront Physics and Mathematics Program to Drive Transformation (FoPM), a World-leading Innovative Graduate Study (WINGS) Program, the University of Tokyo [X.W.], by JSPS Grant-in-Aid for Scientific Research Number 26KJ0757 [X.W.] and 23K20843 [M.Y.], by the Fundamental Research Funds for the Central Universities, and by the Project 12475060 supported by NSFC, Project 24ZR1472400 sponsored by Natural Science Foundation of Shanghai, and Shanghai Pujiang
Program 24PJA134 [Y.Z.]. Kavli IPMU is supported by the World Premier International Research Center Initiative (WPI), MEXT, Japan. X.W. also acknowledges support by grant 63132 from the John Templeton Foundation, as recipient of an Enrico Fermi Fellowship awarded through the Center for SpaceTime and the Quantum. The opinions expressed in this publication are those of the author(s) and do not necessarily reflect the views of the respective funding body.

\section*{Data Availability}
The data analyzed in this study are publicly available in \cite{ding_2026_21128751}.

\appendix

\section{Uniqueness of the Discretized Inverse Problem}\label{app:well_posedness}

To prove the uniqueness of the inverse problem in Eq.~\eqref{eq:probability_1} based on discretized treatment, we consider it in a physical scenario where the detected PBH binaries come from a redshift range $(z_{\min}, z_{\rm max})$ and PBH masses are in the mass range $(m_{\rm min}, m_{\rm max})$. In these parameter ranges, PBH mass function $f_p(m)$, $\eta(m_1,m_2)$, window function $W(m_1, m_2;z)$, and redshift distribution $p(z)$ are positive and well-behaved. Then we can rewrite Eq.~\eqref{eq:probability_1} as
\begin{align}\label{eq:unique_eq}
    P_O(m_1^z, m_2^z) = \int_{z_{\rm min}}^{z_{\rm max}} f_p\left(\frac{m_1^z}{1+z}\right) f_p\left(\frac{m_2^z}{1+z}\right) \eta\left(\frac{m_1^z}{1+z}, \frac{m_2^z}{1+z}\right) W\left(\frac{m_1^z}{1+z}, \frac{m_2^z}{1+z}; z\right) \frac{p(z)}{(1+z)^2} \, dz~.
\end{align}
%
In this proof, We understand it from the data reconstruction perspective by discretizing PBH mass function and kernel function and then Eq.~\ref{eq:unique_eq} can be transferred to a matrix equation.

To simplify our consideration, we set $m_1^z = m_2^z = m_z$ and apply a variable transformation $\frac{m_z}{1+z} = m$, then we can express Eq.~\ref{eq:unique_eq} as
\begin{align}\label{eq:equal_mass}
    P_O(m_z,m_z) = \int_{m_z/(1+z_{\rm max})}^{m_z/(1+z_{\rm min})} f_p\left(m\right) f_p\left(m\right) \eta\left(m, m\right) W\left(m, m; \frac{m_z}{m}-1\right) p\left(\frac{m_z}{m}-1\right)\frac{1}{m_z} \, dm~.
\end{align}
Here we redefine $g(m) \equiv f_p(m)^2$ and $K(m,m_z) \equiv \eta\left(m, m\right) W\left(m, m; \frac{m_z}{m}-1\right) p\left(\frac{m_z}{m}-1\right)\frac{1}{m_z}$ and Eq.~\eqref{eq:equal_mass} reads
\begin{align}\label{eq:kernel}
    P_O(m_z,m_z) = \int_{m_z/(1+z_{\rm max})}^{m_z/(1+z_{\rm min})} g(m) K(m,m_z)dm~.
\end{align}
Then we can discretize Eq.~\eqref{eq:kernel} on the $N\times N$ lattice in $(m,m_z)$ space, and it can be expressed in the form of
\begin{align}
    P_O(m_{z,i},m_{z,i}) = \sum_{j=1}^N w_j K(m_j,m_{z,i}) g(m_j) ~,
\end{align}
where $m_{z,i}$ and $m_j$ is the value of $m_z$ and $m$ at $i$th and $j$th nodes on this $N\times N$ lattice, $w_j$ is the weight of the quadrature formula.
This discrete form can be written in a matrix equation as
\begin{align}
    \mathbf{P} = \mathbf{K} \mathbf{g}~.
\end{align}
Here, the component of vector and kernel matrix follow $\mathbf{P}_i = P_O(m_{z,i},m_{z,i})$, $\mathbf{g}_j = g(m_j)$, and $\mathbf{K}_{ij} = w_j K(m_j,m_{z,i})$. Then the vector $\mathbf{g}$ can be found via an inverse kernel matrix $\mathbf{K^{-1}}$ as
\begin{align}
    \mathbf{g} = \mathbf{K^{-1}} \mathbf{P}~.
\end{align}
The vector solution of $f_p(m)$ can be further expressed as $\mathbf{f}_i = \sqrt{\mathbf{g}_i}$. Since the inverse kernel matrix $\mathbf{K^{-1}}$ is unique, and this promises a unique solution of $\mathbf{g}$. Meanwhile, PBH mass function should be positive, this makes sure a unique vector $\mathbf{f}$ from the vector $\mathbf{g}$. 

However, the noise in the observable vector $\mathbf{P}$ would cause an amplified noise error in $\mathbf{g}$ via the inverse kernel matrix $\mathbf{K^{-1}}$. To avoid such an error in the inverse problem, we introduce the regularization for the inverse problem by solving a minimization problem as
\begin{align}
    \mathbf{g}_\lambda
    =
    \arg\min_{\mathbf{g}_i>0}
    \left[
    \|\mathbf{Kg}-\mathbf{P}\|_2^2
    +
    \lambda \|\mathbf{L}\mathbf{g}\|_2^2
    \right],
\end{align}
and the solution of $\mathbf{g}$ is
\begin{align}
    \mathbf{g} = (\mathbf{K}^T\mathbf{K} + \lambda \mathbf{L}^T \mathbf{L})^{-1} \mathbf{K}^T \mathbf{P}~,
\end{align}
where $\mathbf{L}$ is the regularization operator such as Tikhonov regularization. The additional regularization term $\lambda \mathbf{L}^T \mathbf{L}$ ensures the resulting matrix $\mathbf{K}^T\mathbf{K} + \lambda \mathbf{L}^T \mathbf{L}$ strictly positive definite, and this leads to the existence and uniqueness of the inverse matrix $(\mathbf{K}^T\mathbf{K} + \lambda \mathbf{L}^T \mathbf{L})^{-1}$. Hence, the solution for $\mathbf{g}$ and corresponding PBH mass vector $\mathbf{f}$ exist and are unique for a fixed regularization operator $\mathbf{L}$ and fixed hyperparameter $\lambda$ as what we discussed in Appendix~\ref{app:regularization}. 

\section{Regularization}
\label{app:regularization}
\begin{figure}[htbp]
	\centering
    \includegraphics[width=.43\textwidth]{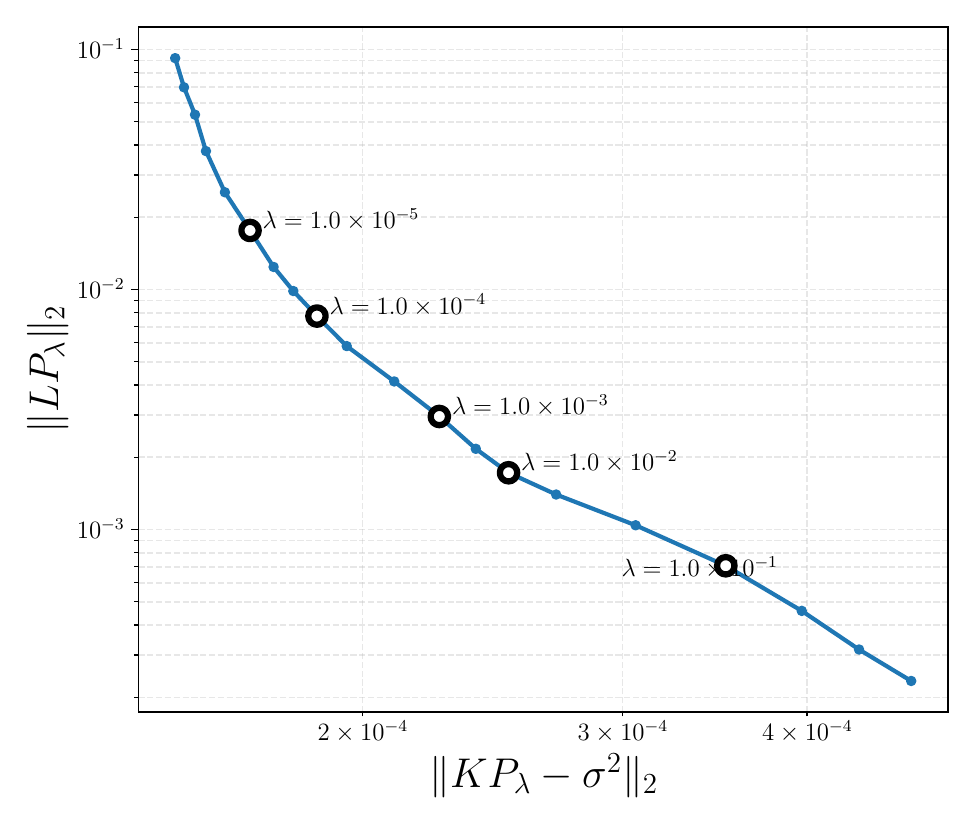}
    \includegraphics[width=.43\textwidth]{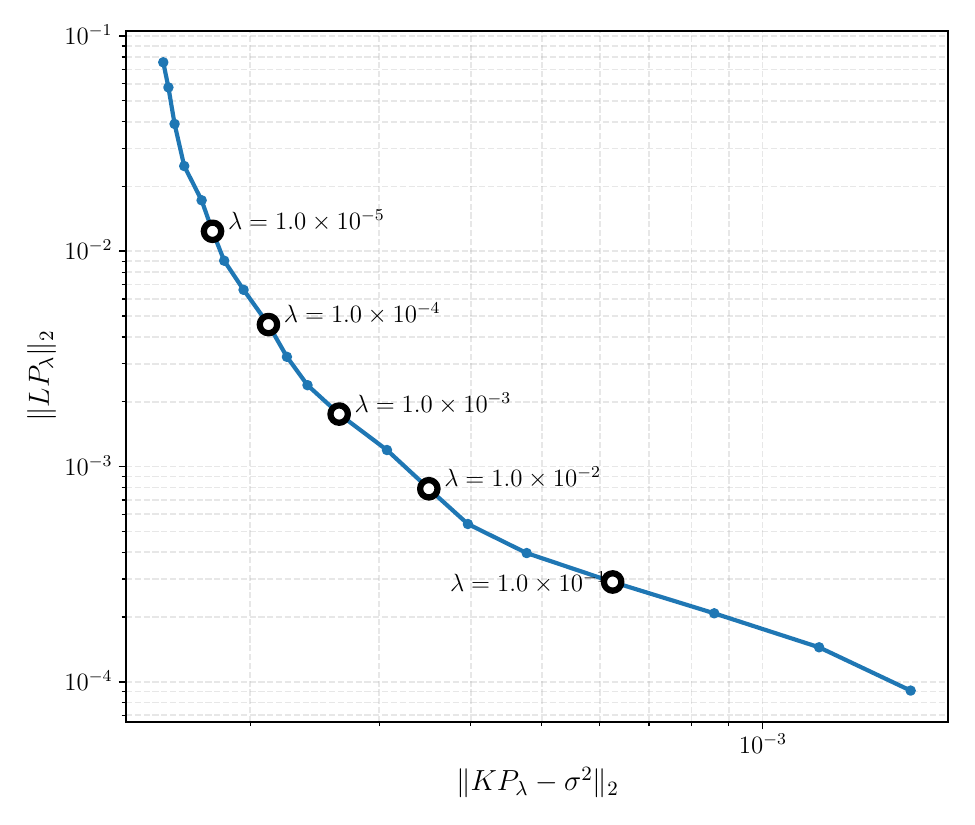}
	\caption{L-curve of the reconstruction (Left: second-order regularization, right: first-order regularization).}
	\label{fig:PPS_Lcurve}
\end{figure}

The choice of \(\lambda\) is guided by the L-curve criterion. 
For each \(\lambda\), we compare the data-misfit norm 
$\rho(\lambda)=\|\mathbf{K}\mathbf{P}_{\lambda}-\boldsymbol{\sigma}^2\|_2$
with the regularization norm 
$\eta(\lambda)=\|\mathbf{L}\mathbf{P}_{\lambda}\|_2$ (See in Fig. \ref{fig:PPS_Lcurve}). 
Small \(\lambda\) values reproduce the input variance more closely but may amplify numerical noise and generate oscillatory spectra, whereas large $\lambda$ values suppress oscillations at the cost of over-smoothing the reconstruction and increasing the residual. 
We therefore focus on values around the bending region of the L-curve, where the balance between data fidelity and smoothness is optimized. 
In the present analysis, we use $\lambda=10^{-5},10^{-4},10^{-3},10^{-2},10^{-1}$ to cover this transition region and treat the resulting spread as a measure of the systematic uncertainty associated with regularization.
\section{Notation and Conventions}\label{app:symbols}

This appendix describes the parameters in this work.

\clearpage
\onecolumngrid
\begin{longtable}{p{3cm} p{10cm} p{3cm}}
    
    \caption{Description of Parameters} 
    \label{tab:variables} \\
    \toprule
    \textbf{Parameter} & \textbf{Description} & \textbf{Introduced in} \\ 
    \midrule
    \endfirsthead
    
    \multicolumn{3}{c}{{ Table \thetable\ continued from previous page}} \\
    \toprule
    \textbf{Parameter} & \textbf{Description} & \textbf{Introduced in}\\ 
    \midrule
    \endhead

    \midrule
    \multicolumn{3}{r}{{Continued on next page}} \\
    \bottomrule
    \endfoot
    
    \bottomrule
    \endlastfoot
    
        $m$ & Mass of PBHs  & Eq.~\eqref{eq:roadmap}\\
        $m_z$ & Redshifted mass of PBHs & Eq.~\eqref{eq:roadmap}\\
        $z$ & Redshift & Eq.~\eqref{eq:roadmap}\\
        $P(m_1^z,m_2^z)$ & Redshifted mass distribution of PBH binaries  & Eq.~\eqref{eq:redshifted_mass_dist}\\
        $P_O(m_1^z,m_2^z)$ & Observed redshifted mass distribution of PBH binaries  & Eq.~\eqref{eq:roadmap}\\
        $C(m_1^z,m_2^z)$ & Cumulative distribution of redshifted mass of PBH binaries  & Eq.~\eqref{eq:cumulative_dist}\\
        $N_{\rm obs}$ & Observed number of PBH binaries on the past light cone & Eq.~\eqref{eq:redshifted_mass_dist}\\
        $N_{\rm tot}$ & Total number of PBH binaries on the past light cone & Eq.~\eqref{eq:redshifted_mass_dist}\\ 
        $N_z(m_1, m_2)$ & The number of PBH binaries of ($m_1$,$m_2$) on the past light cone & Eq.~\eqref{eq:redshift_distribution}\\
        $dn$ & Comoving number density of PBHs in the mass range $(m,m+dm)$ & Eq.~\eqref{eq:pbh_massfun}\\ 
        $n_{\rm PBH}$ & Total comoving number density of PBHs & Eq.~\eqref{eq:pbh_massfun}\\ 
        $f(m)$ & PBH mass function & Eq.~\eqref{eq:roadmap}\\ 
        $f_p(m)$ & Physical PBH mass function & Eq.~\eqref{eq:probability_1}\\
        $\tilde{f}(m)$ & Assumed PBH mass function & Eq.~\eqref{eq:theoretical_prob}\\
        $\mathbf{f}$ & Discrete PBH mass distribution vector & Eq.~\eqref{eq:error_fun}\\
        $n_b$ & Comoving number density of PBH binaries & Eq.~\eqref{eq:dnb}\\
        $\eta\,(m_1,m_2)$ & PBH binary formation mass weighting factor & Eq.~\eqref{eq:pbh_binary_fraction}\\
        $p(z)$ & Redshift distribution of PBH binaries & Eq.~\eqref{eq:redshift_distribution}\\
        $W(m_1,m_2;z)$ & Detectable window function & Eq.~\eqref{eq:window_function}\\
        $E(\mathbf{f})$ & Error function & Eq.~\eqref{eq:error_fun}\\
        $\gamma$ & learning rate in the gradient-descent method & Eq.~\eqref{eq:recursion}\\
        ${\rm SNR}$ & Signal-to-noise ratio & Eq.~\eqref{eq:SNR}\\
        $p_{\rm astro}$ & Probability of astrophysical events & Above Eq.~\eqref{eq:mass_probability}\\
        $S_n$ & Noise strain of the GW detector & Eq.~\eqref{eq:SNR}\\
        $\tilde{h}(f)$ & Fourier transform of the GW strain & Eq.~\eqref{eq:SNR}\\
        $\mathcal{M}_c$ & Chirp mass of PBH binaries & Eq.~\eqref{eq:GW_strain_Fourier}\\
        $d_L$ & Luminosity distance between PBH binaries and the observer & Eq.~\eqref{eq:GW_strain_Fourier}\\
        $a$ & Semi-major axis of PBH binaries & Eq.~\eqref{eq:orbit_para_prob}\\
        $a_{\rm ini}$ & Initial semi-major axis of PBH binaries & Eq.~\eqref{eq:prob_t}\\
        $a_t$ & Semi-major axis of PBH binaries at cosmic time $t$ & Eq.~\eqref{eq:prob_t}\\
        $e$ & Eccentricity & Eq.~\eqref{eq:orbit_para_prob}\\
        $P_{\rm ini}$ & Initial probability distribution of PBH binary parameters & Eq.~\eqref{eq:orbit_para_prob}\\
        $P_t$ & Probability distribution of PBH binary parameters at cosmic time $t$ & Eq.~\eqref{eq:prob_t}\\
        $\Theta$ & Heaviside step function & Eq.~\eqref{eq:windowfun}\\
        $f$ & GW frequency & Eq.~\eqref{eq:windowfun}\\
        $f_c$ & Critical GW frequency & Eq.~\eqref{eq:windowfun}\\
        $f_b$ & Energy density fraction of PBH binary in dark matter & Eq.~\eqref{eq:orbit_para_prob}\\
        $\bar{x}$ & Mean separation of PBH binaries & Eq.~\eqref{eq:orbit_para_prob}\\
        $V_c$ & Comoving volume on the past light cone & Eq.~\eqref{eq:redshift_dist_comoving_volume}\\
        $D_H$ & Hubble distance & Below Eq.~\eqref{eq:redshift_dist_comoving_volume}\\
        $D_M$ & Transverse distance & Below Eq.~\eqref{eq:redshift_dist_comoving_volume}\\
        $E(z)$ & Dimensionless Hubble parameter & Below Eq.~\eqref{eq:redshift_dist_comoving_volume}\\
        $p(m_z)$ & Probability distribution of redshifted mass of PBHs & Eq.~\eqref{eq:mass_probability}\\
        $\mu$ & Mean value of probability distribution of redshifted mass & Eq.~\eqref{eq:mass_probability}\\
        $\sigma_{\pm}$ & Standard deviation of probability distribution of redshifted mass & Eq.~\eqref{eq:mass_probability}\\
        $m_c$ & Characteristic PBH mass & Eq.~\eqref{eq:LN_PL}\\
        $\sigma_{\mathrm{MF}}$ & Width of PBH mass function & Eq.~\eqref{eq:LN_PL}\\
        $f_{\rm LN}(m)$ & Lognormal PBH mass function & Eq.~\eqref{eq:LN_PL}\\
        $\alpha_{\mathrm{MF}}$ & Power index of power-law PBH mass function & Eq.~\eqref{eq:LN_PL}\\
        $A_{\mathrm{MF}}$ & Dimensionless factor in power-law PBH mass function & Eq.~\eqref{eq:LN_PL}\\
        $f_{\rm PL}(m)$ & Power-law PBH mass function & Eq.~\eqref{eq:LN_PL}\\
        $\chi_\nu^2$ & Reduced chi-square & Eq.~\eqref{eq:reduced_chi_square}\\
        $R$ & Volumetric merger rate density & Eq.~\eqref{eq:merger_rate_density}\\
        $t(z)$ & Cosmic time at redshift $z$ & Eq.~\eqref{eq:merger_rate_density}\\
        $t_0$ & Cosmic time at redshift $z = 0$ & Eq.~\eqref{eq:merger_rate_density}\\
        $f_{\rm PBH}$ & Energy density fraction of PBHs in dark matter & Eq.~\eqref{eq:merger_rate_density}\\
        $S(f_{\rm PBH})$ & Suppression factor in PBH merger rate & Eq.~\eqref{eq:merger_rate_density}\\
        $\langle m \rangle$ & Average mass of PBHs & Eq.~\eqref{eq:merger_rate_density}\\
        $T_{\rm obs}$ & Run duration time of LVK & Eq.~\eqref{eq:observed_pbh_num}\\
        $N_{\rm PBH}$ & The number of detectable PBH merger events & Eq.~\eqref{eq:observed_pbh_num}\\
        $\delta$ & Density perturbation & Eq.~\eqref{eq:delta_R}\\
        $\delta_c$ & Threshold value of density perturbation to form PBHs & Above Eq.~\eqref{eq:delta_R}\\
        $R$ & Comoving scale R & Eq.~\eqref{eq:delta_R}\\
        $k$ & Comoving wavenumber $k$ & Eq.~\eqref{eq:sigma_square}\\
        $W(R;x-x')$ & Window function for density perturbation on scale $R$ & Eq.~\eqref{eq:delta_R}\\
        $\sigma^2(R)$ & Variance of density perturbation after smoothing & Eq.~\eqref{eq:sigma_square}\\
        $\mathcal{P}_\delta$ & Power spectrum of the density perturbation & Eq.~\eqref{eq:P_delta}\\
        $\mathcal{P}_{\mathcal{R}}$ & Power spectrum of comoving curvature perturbation & Eq.~\eqref{eq:P_delta}\\
        $w$ & Equation of state of the epoch during PBH formation & Eq.~\eqref{eq:P_delta}\\
        $\tilde{W}(k,R)$ & Volume-normalized Gaussian window function & Eq.~\eqref{eq:sigma_square}\\
        $m(R)$ & Mass of PBHs from comoving scale $R$ & Eq.~\eqref{eq:m_R}\\
        $M_H$ & Hubble horizon mass & Eq.~\eqref{eq:m_R}\\
        $\gamma_m$ & PBH collapse efficiency factor & Eq.~\eqref{eq:m_R}\\
        $g_\star$ & The number of relativistic degrees of freedom & Eq.~\eqref{eq:m_R}\\
        $\beta$ & PBH abundance & Eq.~\eqref{betasigma}\\
        $f_{\rm PBH} (m)$ & PBH fraction mass function in dark matter & Eq.~\eqref{eq:fpbh_m_normalization}\\
        $T_{\rm form}$ & Temperature at PBH formation & Eq.~\eqref{eq:fpbh_m}\\
        $\mathbf{K}$ & Kernel matrix for $K(k,R)$ & Eq.~\eqref{eq:vector_sigma_square}\\
        $\vec{\sigma^2}$ & Discretized vector for $\sigma^2(m)$ & Eq.~\eqref{eq:vector_sigma_square}\\
        $\vec{\mathcal{P}}$ & Discretized vector for primordial power spectrum & Eq.~\eqref{eq:vector_sigma_square}\\
        $\mathbf{L}$ & Regularization matrix & Eq.~\eqref{eq:minimization}\\
        $\lambda$ & Hyperparameter for the smoothness prior & Eq.~\eqref{eq:minimization}\\
        $\mathbf{R}$ & Reconstruction matrix & Eq.~\eqref{eq:solution_P_lambda}\\
        $\sigma_\lambda$ & Uncertainty in the reconstructed spectra & Eq.~\eqref{eq:sigma_lambda}\\
        $\alpha_p$ & Factor of power-law distribution in parameterization of PPS & Eq.~\eqref{eq:fitting_fun_shape}\\
        $n_p$ & Power index of power-law distribution in parameterization of PPS & Eq.~\eqref{eq:fitting_fun_shape}\\
        $\beta_p$ & Factor of lognormal distribution in parameterization of PPS & Eq.~\eqref{eq:fitting_fun_shape}\\
        $\sigma_p$ & Width of lognormal distribution in parameterization of PPS & Eq.~\eqref{eq:fitting_fun_shape}\\
        $k_{\rm peak}$ & Peak scale in parameterization of PPS & Eq.~\eqref{eq:fitting_fun_shape}\\
        $P_G$ & Gaussian part distribution in PPS & Eq.~\eqref{eq:Edgeworth expansion}\\
        $H_n$ & Probabilists' Hermite polynomials & Eq.~\eqref{eq:Edgeworth expansion}\\
        $S_3$ & Skewness & Eq.~\eqref{eq:Edgeworth expansion}\\
        $S_4$ & Kurtosis & Eq.~\eqref{eq:Edgeworth expansion}\\
        $\rho\,(\lambda)$ & Data-misfit norm & Appendix.~\ref{app:regularization}\\
        $\eta\,(\lambda)$ & Regularization norm & Appendix.~\ref{app:regularization}\\
        
\end{longtable}

\bibliographystyle{utphys}
\bibliography{reconstruct}

\end{document}